\documentclass[referee]{gji}
\usepackage[latin1]{inputenc}
\usepackage{amsmath}
\usepackage{amsfonts}
\usepackage{amssymb}
\usepackage{graphicx,epsf,subfigure}
\usepackage{epsfig}
\usepackage{epstopdf}
\usepackage{times}
\usepackage{color}
\let\bibhang\relax
\usepackage{natbib}
\usepackage{multirow}

\begin{document}
\label{firstpage}
\author[S. Vantieghem, A. Sheyko and A. Jackson]{S. Vantieghem, A. Sheyko, A. Jackson \\ Institut f\"ur Geophysik, ETH Z\"urich, Sonneggstrasse 5, CH8092 Z\"urich, Switzerland}
\title{Applications of a  finite-volume algorithm for incompressible MHD problems}
\maketitle
\begin{abstract}
We present the theory, algorithms and implementation of a parallel finite-volume algorithm for the solution of the incompressible magnetohydrodynamic (MHD) equations using unstructured grids that are applicable for a wide variety of geometries.
Our method implements a mixed Adams-Bashforth/Crank-Nicolson scheme for
the nonlinear terms in the MHD equations and we prove that it is stable independent of the time step. To ensure that the solenoidal condition is met for the magnetic field, we
use a method whereby a pseudo-pressure is introduced into the induction equation; since we are concerned with incompressible flows, the resulting Poisson equation for the pseudo-pressure is solved alongside the equivalent Poisson problem for the velocity field.
We validate our code in a variety of geometries including periodic boxes, spheres, spherical shells, spheroids and ellipsoids; for the finite geometries we implement the so-called ferromagnetic or pseudo-vacuum boundary conditions appropriate for a surrounding medium with infinite magnetic permeability. This implies that the magnetic field must be purely perpendicular to the boundary. We present a number of comparisons against previous results and against analytical solutions, which verify  the code's accuracy. This documents the code's reliability as a prelude to its use in more difficult problems.
We finally present a new simple drifting solution for thermal convection in a spherical shell that successfully sustains a
magnetic field of simple geometry. By dint of its rapid stabilization from the given initial conditions, we deem it suitable as a benchmark against which other self-consistent dynamo codes can be tested.
\end{abstract}  
\section{Introduction}
The Earth, as well as many other celestial bodies, posesses a dynamo, that is, a self-sustained magnetic field that is generated by the turbulent flow of molten iron-rich material in its fluid outer core. Understanding the processes governing the geodynamo, and more broadly, the dynamics of planetary fluid envelopes, is a formidable task for numerous reasons. First, we can only probe the interiors of these bodies indirectly, for example through observations of the magnetic field at the planetary surface, analysis of seismic data, records of length-of-day variations or detailed measurements of their gravity field. Moreover, the timescales involved range from days to millions of years whereas abundant data coverage has been available only for  a few decades. Furthermore, planetary cores operate in an extreme parameter regime that is difficult to reach experimentally. Bearing all this in mind, numerical simulations of the relevant core processes are essential to increase our understanding of natural dynamo action.

The mechanisms underlying the generation and sustenance of magnetic fields are studied in a branch of physics called magnetohydrodynamics (MHD). The notion MHD refers to physical processes in which there is a mutual interaction between the flow of an electrically conducting fluid and an electromagnetic field. In mathematical terms, this leads to a coupling between the Navier\-Stokes equations and the pre-Maxell equations. Apart from geo- and astrophysical phenomena, MHD flows are also encountered in metallurgical processes where magnetic fields are used for various purposes such as the enhancing of mixing processes or the damping of small-scale turbulence. The essential difference between natural and manmade MHD phenomena is that the coupling between flow and magnetic field is virtually one-way in the latter. More specifically, in manmade flows, the role of the Lorentz force in the momentum balance can be significant, whereas the magnetic field induced by the flow mostly remains negligible with respect to the externally imposed one. This difference can be cast in terms of the magnetic Reynolds number $Re_m$, as defined later by equation (\ref{eq:def_Rem}), which is a proxy for the ability of the flow to induce a significant magnetic field. This implies that the existence of a dynamo demands that the magnetic Reynolds number is large. 

Starting with the pioneering work of \citet{Glatzmaier1995} and \citet{kageyama1995}, numerical studies of self-consistent geodynamo processes in a planetary geometry have mainly been performed using spectral codes. These methods are based on the following strategy: (1) The divergence-free character of the velocity and magnetic field is implicitly built-in by using a toroidal-poloidal formulation. (2) The unknown scalar functions are expanded in a complete series of functions. In the spherical case spherical harmonics and a set of radial basis functions are used; the 1-D problem in the radial direction is then solved by means of a finite-difference approach or a spectral method that uses basis functions such as the Chebychev or Jacobi polynomials. This type of code is nowadays well-established, and the continuous increase in computational power has allowed more Earth-like parameters \citep[e.g.,][]{Sakuraba2009} to be reached. The main limitation of these codes is that their application is restricted to problems with a spherical symmetry; a small drawback is the computational expense from the back-and-forth transformations between spectral and real space required for the evaluation of nonlinear terms; this brings along couplings between all expansion coefficients and global communication in simulations using distributed memory parallelization.

Local discretization approaches like the finite-volume (FV), finite-difference or finite-element method have been much less popular for the simulation of dynamo processes. The rationale for using this type of code is that they can accommodate more easily boundary topography, which has recently been advocated to be an essential ingredient in the mechanism underlying the dynamos of the ancient Moon and exoplanets \citep{dwyer2011,lebarsNature,cebronAA}. This type of code essentially faces two challenging problems: 1) The discretized induction equation does not necessarily conserve the solenoidal character of the magnetic field. 2) The boundary condition next to an electrically insulating medium, representing the silicate mantle in the context of planetary physics, has a non-local nature, and requires, in principle, the solution of a Laplace equation in an exterior domain of infinite extent. The first issue can be avoided by using a so-called constrained transport method \citep[][]{evans1988,teyssier2006}, which is based on the rotational form of the induction equation (see expression (\ref{eq:induction_rot}) below). To date, this technique has only been used in the context of structured-grid codes (albeit in a non-Cartesian coordinate system). Some other authors have reformulated the induction equation in terms of the magnetic vector potential \citep{matsui2004development,CebronMHD}. A third possibility is to supplement the induction equation with the gradient of a pseudo-pressure that acts as a Lagrangian multiplier to project the magnetic field onto a solenoidal field. This approach was introduced by \cite{toth2000}, and has since been adopted by a large number of authors \citep[e.g.,][]{harder2005finite,chan2007,guermond2007}. 

The issue of insulating boundary conditions has been approached from different angles as well. Given the fact that, in the insulating exterior, the magnetic field derives from a scalar potential $\phi$ which declines as $\mathcal{O}(r^{-2})$, one can solve the Laplace equation governing $\phi$ in an extended but finite exterior domain and impose $\phi=0$ far away from the fluid domain of interest. Another approximation is to replace the insulating exterior by a weakly conducting one \citep{chan2007}. An elegant alternative was devised by \citet{iskakov2004}, who recasted the Laplace equation for $\phi$ into a boundary integral equation on the interface between the insulating and conducting domain. This method does not introduce any approximation at the physical level, and reduces a 3-D problem into a 2-D one. The main disadvantage of this method, however, is that the coefficient matrix representing the discretized boundary integral equation is dense so that this approach eventually comes at a higher computational cost than the direct solution of the Laplace equation for $\phi$. Apart from the computational overhead associated with solving the Laplace equation, the different approximations discussed above also affect the accuracy of the results obtained with these codes, as noted by \citet{jackson2014}. 

Pseudo-vacuum boundary conditions, also referred to as ferromagnetic boundary conditions, have become a popular alternative to the use of insulating boundary conditions. Implemented for the first time in a numerical code by \citet{kageyama1995}, these conditions prescribe that the magnetic field tangential to the boundary of the conducting fluid region to be zero. The physical equivalent of this mathematical condition is a perfect ferromagnetic exterior, that is, a material whose magnetic permeability tends to infinity. These conditions are more easily implemented in local codes, and do not require additional approximations. Hence, they may be better suited for benchmarking purposes. Finally, this type of boundary condition has also been applied by the solar physics community, where its use is underpinned by observational support \citep{gilmanmiller}.

In this work, we present a FV algorithm for the solution of the incompressible MHD equations. It is akin to the ones used by \citet{harder2005finite,hullermann} and \cite{wu2009dynamo,WR2012}. The main novelty of our algorithm is the capability of handling arbitrary unstructured meshes. This permits us to consider a wide range of different geometries, like spheres and ellipsoids both with and without inner cores. To describe our approach, we first present the governing equations in section \ref{sec:eqns}. Then, we describe and validate our numerical method in sections \ref{sec:bulk_algorithm} and \ref{sec:ferro}. In section \ref{sec:benchmark}, we present a self-consistent numerical dynamo benchmark. 

\section{Mathematical background} \label{sec:eqns}
The equations governing incompressible MHD can be readily derived from the incompressible Navier-Stokes equation and the pre-Maxell equations. From a physical point of view, the neglect of displacement currents is valid as long as the flow speed is small with respect to the speed of light in the medium, that is, for non-relativistic phenomena. Upon the choice of a characteristic length and velocity scale $L$ and $U$, respectively, we can write the MHD equations in the following non-dimensional form:
\begin{eqnarray}
\nabla \cdot {\boldsymbol u} & = & 0, \label{eq:divu} \\
\frac{\partial {\boldsymbol u}}{\partial t} + {\boldsymbol u} \cdot \nabla {\boldsymbol u}  & = & {\boldsymbol b} \cdot \nabla {\boldsymbol b} + Re^{-1}\nabla^2{\boldsymbol u} - \nabla p + {\boldsymbol f}_b, \label{eq:navierstokes} \\
\frac{\partial {\boldsymbol b}}{\partial t} + {\boldsymbol u} \cdot  \nabla {\boldsymbol b} & = & {\boldsymbol b}\cdot \nabla {\boldsymbol u} + Re_m^{-1} \nabla^2{\boldsymbol b}, \label{eq:induction} \\
\nabla \cdot {\boldsymbol b} & = & 0. \label{eq:divb}
\end{eqnarray}
Here $t$ denotes time, ${\boldsymbol u}$ the velocity field, ${\boldsymbol b}$ the magnetic field, ${\boldsymbol f}_b$ a body force, and $p$ the (modified) pressure. These quantities have been non-dimensionalised according to ${\boldsymbol u} \rightarrow U{\boldsymbol u}$, $t \rightarrow LU^{-1} t$, $\nabla \rightarrow L^{-1} \nabla$, $p \rightarrow \rho U^{2} p$, ${\boldsymbol f}_b \rightarrow \rho U^2 L^{-1}{\boldsymbol f}_b $, ${\boldsymbol b} \rightarrow U \sqrt{\rho \mu}{\boldsymbol b}$, where $\rho$ and $\mu$ denote the fluid's mass density and magnetic permeability, respectively.
 
These equations contain two independent non-dimensional numbers: the well-known Reynolds number $Re$,
\begin{equation} 
Re = \frac{UL}{\nu},
\end{equation}
and  the magnetic Reynolds number $Re_m$, \label{eq:def:Re}
\begin{equation}
Re_m = \mu \sigma U L. \label{eq:def_Rem}
\end{equation} 
In these definitions, the symbols $\nu$ and $\sigma$ represent the fluid's viscosity and electrical conductivity, respectively; these material properties will be assumed constant throughout this work.

The induction equation (\ref{eq:induction}) has an alternative representation,
\begin{equation}
\frac{\partial {\boldsymbol b}}{\partial t} = \nabla \times ({\boldsymbol u} \times {\boldsymbol b} - Re_m^{-1} \nabla \times {\boldsymbol b}), \label{eq:induction_rot}
\end{equation}
which reveals immediately that the magnetic induction equation (\ref{eq:induction}) or (\ref{eq:induction_rot}) satisfies the property:
\begin{equation}
\frac{\partial( \nabla \cdot {\boldsymbol b})}{\partial t} = 0.
\end{equation}
This implies that equations (\ref{eq:induction}) or (\ref{eq:induction_rot}) will satisfy the solenoidality constraint (\ref{eq:divb}) for $t>t_0$, provided that $\nabla \cdot {\boldsymbol b}=0$ at $t=t_0$. This property, however, does not necessarily carry over to spatial discretizations of the induction equation. Therefore, special care needs to be taken to ensure that $\nabla \cdot {\boldsymbol b}=0$ when solving numerical approximations of the induction equation.

In order to close the system of equations (\ref{eq:divu})-(\ref{eq:divb}), we should supply suitable boundary conditions for ${\boldsymbol u}$, ${\boldsymbol b}$ at the boundaries of the solution domain $V$, denoted as $\partial V$. Furthermore, we will use the notation $\hat {\boldsymbol n}$ for the external unit normal on $\partial V$. For the velocity field ${\boldsymbol u}$, we will only use no-slip conditions, that is, ${\boldsymbol u}|_{\partial V}=0$. For the magnetic field ${\boldsymbol b}$, we will use the pseudo-vacuum condition, which specifies that the tangential components of ${\boldsymbol b}$ vanish at the domain boundary, that is:
\begin{equation}
\left. {\boldsymbol b} \times \hat{\boldsymbol n}\right|_{\partial V} = 0. \label{eq:bcb}
\end{equation}
It should be noted that this condition does not explicitly prescribe a condition on the wall-normal component of the magnetic field. Indeed, the specific nature of the induction equation does not allow imposition of constraints on the three components of the magnetic field independently, because this could possibly lead to a violation of the solenoidal character of ${\boldsymbol b}$. Rather, the divergence constraint (\ref{eq:divb}) together with the two scalar conditions (\ref{eq:bcb}) will implicitly impose a condition on the boundary-normal component. 

\section{Numerical method} \label{sec:bulk_algorithm}
In this section, we present the implementation of an unstructured FV algorithm for the solutions of the incompressible MHD equations (\ref{eq:divu})-(\ref{eq:divb}), which builds upon the earlier work of \cite{vantieghem_phd} that concerned numerical solutions of the quasi-static MHD equations. 
The basic building block of the FV method is a small control volume (CV), into which the solution domain is divided.  FV methods solve a weak form of the equations under consideration by taking their integral over the CV, and transforming volume integrals of spatial derivatives into surface integrals  over the surface bounding the CV by virtue of Gauss's theorem. The surface integrals are then discretized using Taylor expansions, and this eventually gives rise to a set of ordinary differential equations (ODEs). To describe in more detail this process, first we explain the grid arrangement and how the CVs are constructed. Then, we present the discretization stencils followed by a discussion of how the resulting ODEs are integrated in time. Finally, we validate our method by performing nonlinear dynamo simulations in a periodic box and comparing them against results obtained with a spectral code.  

\subsection{Grid arrangement}
\begin{figure}
\begin{center}
\includegraphics[width=0.6\columnwidth]{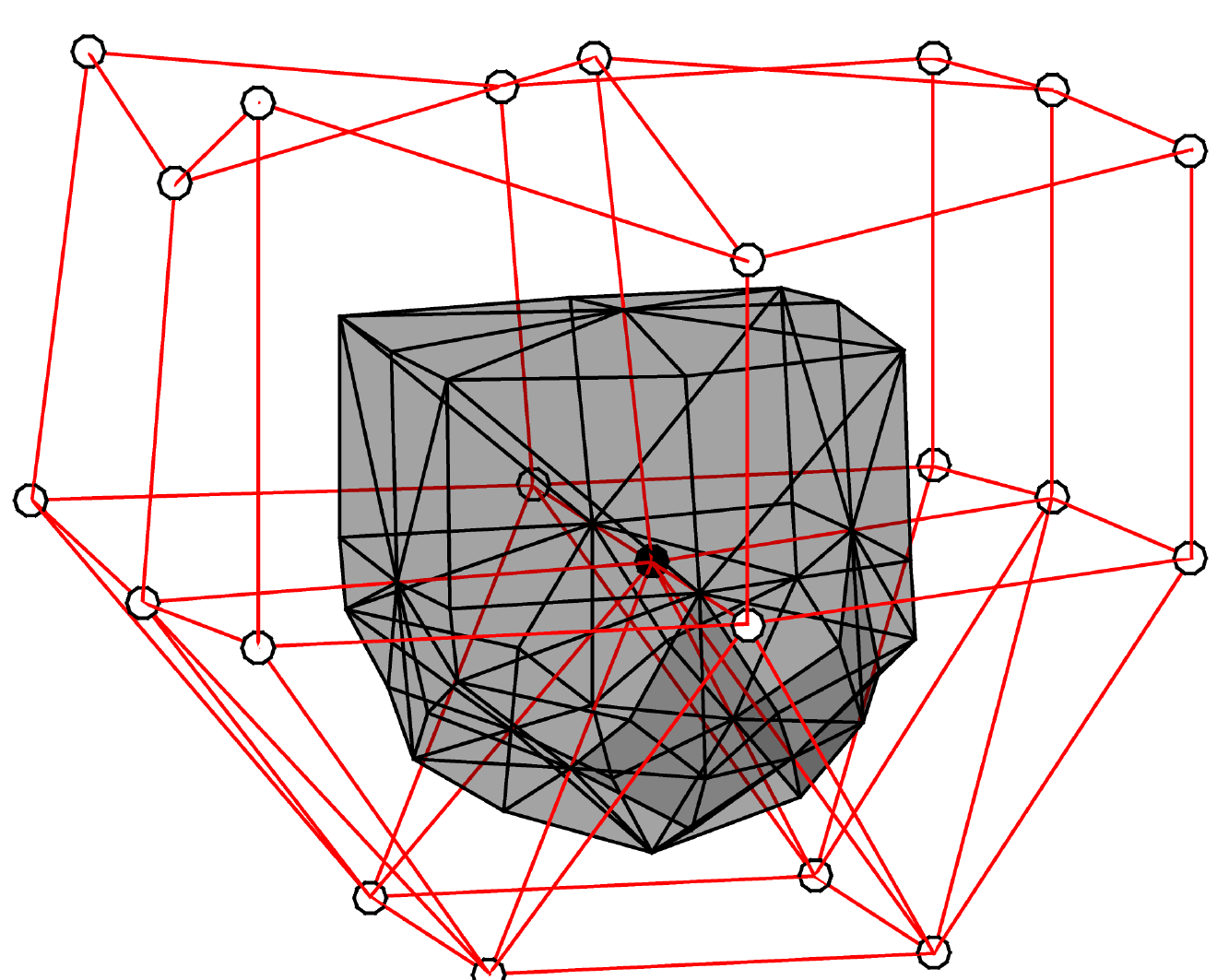}
\caption{Illustration of the construction of a control volume (grey shading) around a grid node (black dot) that is a vertex of an unstructured set of hexahedral, pyramidal and tetrahedral grid elements (red lines).}
\label{fig:cv_total}
\end{center}
\end{figure}

\begin{figure}             
  \begin{center}
  \setlength{\epsfysize}{6.0cm}
    \begin{tabular}{ccc}
      \setlength{\epsfysize}{6.0cm}
      \subfigure[]{\epsfbox{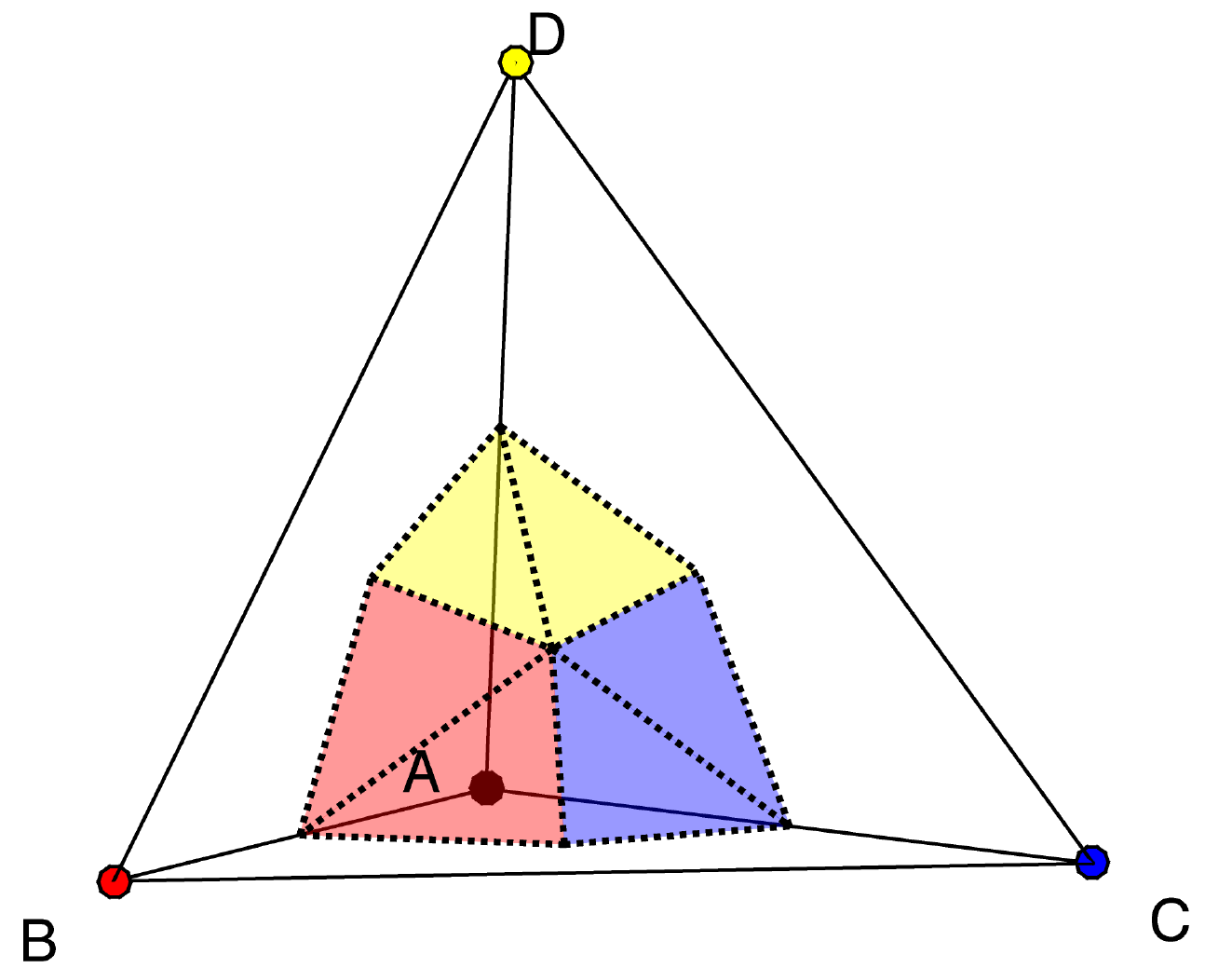}}  &
      \setlength{\epsfysize}{6.0cm}
      \subfigure[]{\epsfbox{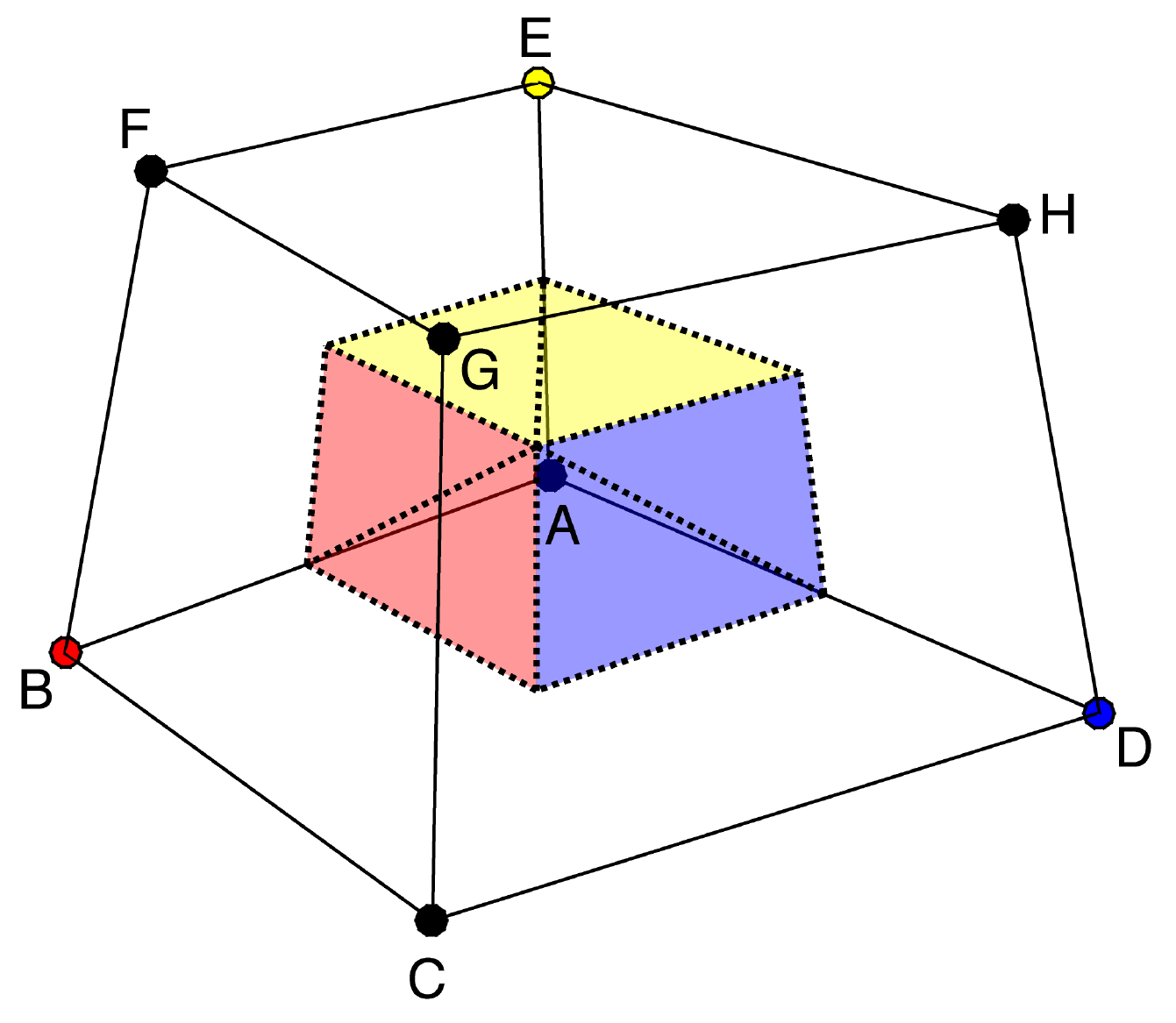}}
    \end{tabular}
    \caption{Association of control volume surface patches to a pair of nodes for (a) a tetrahedral and (b) a hexahedral element. The colors of the different surface patches of the CV refer to the pair with which it is associated.}
         \label{fig:cv}                
  \end{center}
\end{figure}

The starting point for our numerical procedure is a number of non-overlapping polyhedra (tetrahedra, prisms, pyramids or hexahedra) that fill completely the solution domain $V$ and together define the computational grid. We will refer to these basic building blocks as elements and their vertices are termed `grid nodes'. The grid can be unstructured, that is, the connectivity between the grid elements can be arbitrarily complex. In this work, we will adopt a node-based approach, which means that control volumes are created around the grid nodes, as illustrated in Figure \ref{fig:cv_total}. More specifically, the surface that bounds the CV associated with a node $i$ is the union of triangular patches that are created by connecting the midpoints of an edge, the centroid of a face that contains the edge, and the centroid of the element. Each of these elementary triangles can be associated with a pair of grid nodes. For instance, the red triangles in Figure \ref{fig:cv} are associated with the pair A and B; two nodes form a pair if and only if they share a common edge. We will denote by $\pi_i$ the set of all nodes $j$ with which node $i$ forms a pair. Furthermore, the notation ${\boldsymbol S}_{ij}$ refers to the sum of the surface normals of all triangular patches associated with the pair $(i,j)$, whereby the magnitude of a surface normal on a triangular patch is the area of the patch. We adopt the convention that ${\boldsymbol S}_{ij}$ points from $i$ to $j$.
Finally, we will use the notation $V_{i}$ for the volume of the CV associated with node $i$ and $\partial V_i$ for its bounding surface.

We choose a so-called collocated approach, which means that all variables (${\boldsymbol u}$, ${\boldsymbol b}$, $p$,...) share the same CVs. This is preferred above a staggered approach as we avoid the computational complexity and overhead associated with having separate sets of CVs for different variables. In the area of classical hydrodynamical computational fluid dynamics, it is well known that the use of a collocated approach also requires the definition of quantities at the interfaces between CVs (see, e.g. \cite{ferzigerbook}). In the case of incompressible MHD, we will need the discretized equivalent of the momentum and magnetic fluxes through the surface patches ${\boldsymbol S}_{ij}$, that is, $\int_{S_{ij}} {\boldsymbol u} \cdot \mathrm{d}{\boldsymbol S}$ and $\int_{S_{ij}} {\boldsymbol b} \cdot \mathrm{d}{\boldsymbol S}$; in our non-dimensionalized setup we will denote these quantities as ${U}_{ij}$ and $B_{ij}$, respectively.   

\begin{figure}                 
  \begin{center}
\includegraphics[width=0.6\columnwidth]{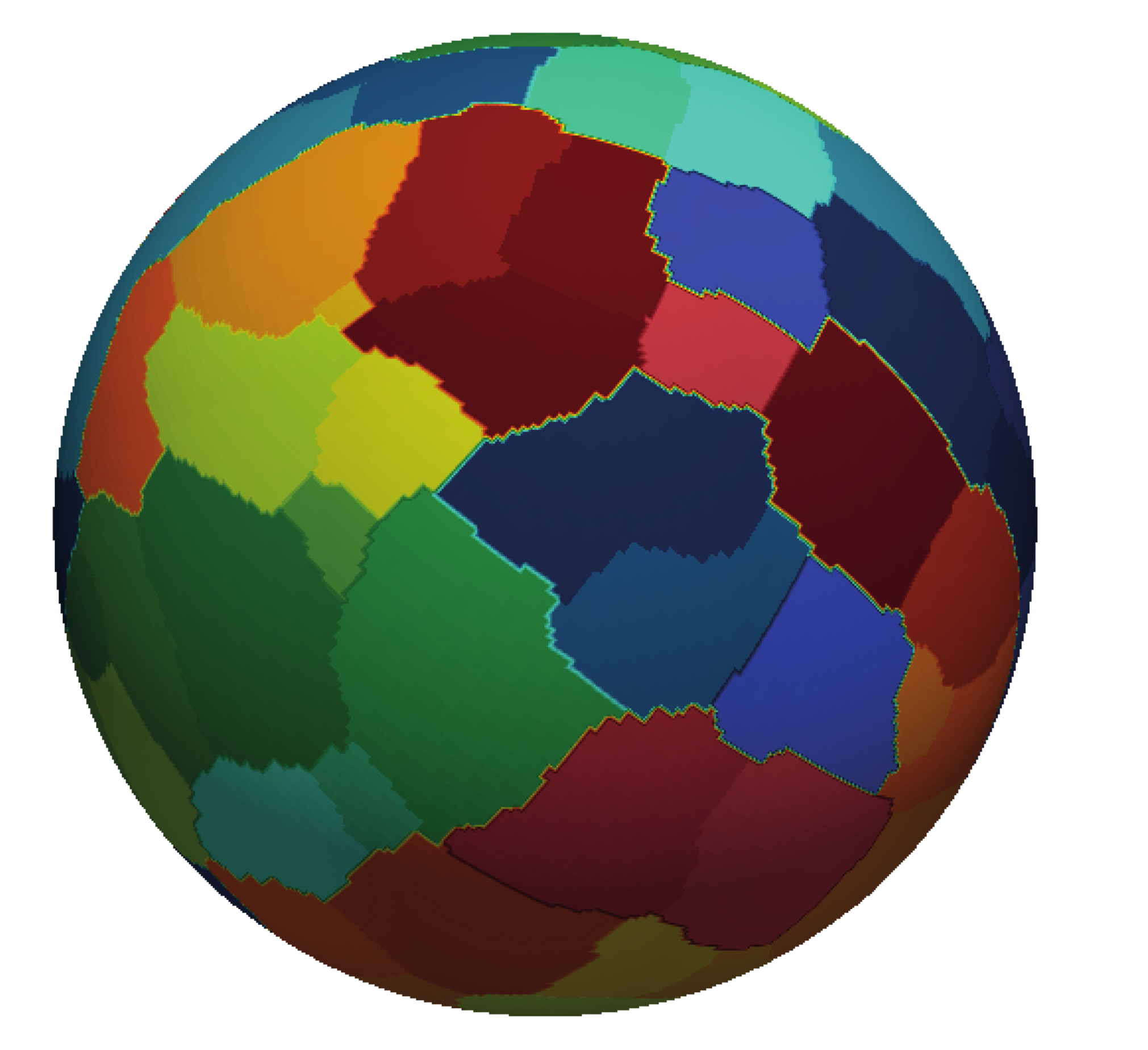}
    \caption{Illustration of the domain decomposition onto different cores of an unstructured grid. The different colours indicate different partitions; please note that some colours have been reused.}
         \label{fig:partition}                
  \end{center}
\end{figure}

The code runs in parallel on distributed-memory systems. The parallelization strategy is based on a domain decomposition approach using the graph partitioning tool \texttt{Metis} \citep{parmetis}. An illustration of the decomposition of a spherical domain is given in Figure \ref{fig:partition}, where each colour corresponds to the part of the grid that is associated with one core. There is in principle no constraint on the number of cores that can be used.

\subsection{Spatial discretization} \label{subsec:spatial}
The essential idea underlying the FV method is that integrals of spatial derivatives over a CV can be transformed into surface integrals by virtue of Gauss's theorem. For instance, the numerical approximation for the divergence of a (continuous) vector field ${\boldsymbol f}$ is the following:
\begin{eqnarray}
\int_{V} \nabla \cdot {\boldsymbol f} \,\mathrm{dV} & = & \oint_{\partial V} {\boldsymbol f} \cdot {\mathrm{d}}{\boldsymbol S} \\ & \approx &  \sum_{j \in \pi_i} {\boldsymbol f} \left( \frac{{\boldsymbol r}_i + {\boldsymbol r}_j}{2}\right) \cdot {\boldsymbol S}_{ij} \\ &  \approx & \sum_{j \in \pi_i} \frac{{\boldsymbol f}({\boldsymbol r}_i) + {\boldsymbol f}({\boldsymbol r}_j)}{2} \cdot {\boldsymbol S}_{ij},
\end{eqnarray}
We now denote by ${\boldsymbol F}$ the vector that contains the values of ${\boldsymbol f}$ at all grid nodes $i$, that is, ${\boldsymbol F}_{i} = {\boldsymbol f}({\boldsymbol r}_i)$ and by ${\mathbb D}$ the discretized divergence operator. In the remainder of this section, we will adopt the convention that capital letters denote spatially discrete variables whereas small letters represent continuous fields. Thus, we have:
\begin{equation}
\left({\mathbb D} {\boldsymbol F} \right)_i = \frac{1}{V_i}\sum_{j \in \pi_i} \frac{\boldsymbol F_i + \boldsymbol F_j}{2} \cdot {\boldsymbol S}_{ij}.  \label{eq:div_discrete}
\end{equation}
Similarly, for the gradient of a scalar function $\phi$, we obtain
\begin{eqnarray}
\int_{V} \nabla \phi\,\mathrm{d}V  & = & \oint_{\partial V} {\phi}\,{\mathrm{d}}{\boldsymbol S} ,\\ & \approx & \sum_{j \in \pi_i} \frac{\phi({\boldsymbol r}_i) +  \phi({\boldsymbol r}_j)}{2}  {\boldsymbol S}_{ij}. 
\end{eqnarray}
We write, using an obvious notation:
\begin{equation}
(\mathbb{G} \Phi)_{i} = \frac{1}{V_i}\sum_{j \in \pi_i} \frac{ \Phi_i + \Phi_j}{2}  {\boldsymbol S}_{ij}. \label{eq:grad_discrete}
\end{equation}
The MHD equations also contain convective-like derivatives of the form ${\boldsymbol f} \cdot \nabla {\boldsymbol g}$, where ${\boldsymbol f}$ and ${\boldsymbol g}$ are solenoidal vector fields. These derivatives can be approximated as
 \begin{eqnarray}
\int_{V} {\boldsymbol f} \cdot \nabla {\boldsymbol g} \,\mathrm{d}V & = & \oint_{\partial V} {\boldsymbol g} \left( {\boldsymbol f} \cdot {\mathrm{d}}{\boldsymbol S} \right), \\ & \approx & \sum_{j \in \pi_i} \frac{{\boldsymbol g}({\boldsymbol r}_i) + {\boldsymbol g}({\boldsymbol r}_j)}{2} F_{ij}.
\end{eqnarray}
We will denote the discretized version of the convective derivative in the following way:
\begin{equation} 
(\mathbb{C}_F {\boldsymbol G})_{i} = \frac{1}{V_i}\sum_{j \in \pi_i} \frac{\boldsymbol G_i + \boldsymbol G_j}{2} F_{ij}. 
 \end{equation}
Furthemore, we also need a discretization scheme for the vector Laplacian operator required for the calculation of the diffusive terms in the momentum and induction equation. Since our code uses a Cartesian coordinate system, this is equivalent to the evaluation of a scalar Laplacian, that is, $\nabla^2 {\boldsymbol u} = (\nabla^2 u_x) \hat{\boldsymbol x} + (\nabla^2 u_y) \hat{\boldsymbol y} + (\nabla^2 u_z) \hat{\boldsymbol z}$.  In addition, the numerical solution of Poisson equations will be required for our time-stepping approach as described in Section \ref{subsec:timestep}.  A straightforward combination of the discretization stencils (\ref{eq:div_discrete}) and (\ref{eq:grad_discrete}) would lead to the well-known odd-even decoupling problem \citep{ferzigerbook}, that is, a stencil that has a non-trivial null space, and is therefore not suitable for the solution of Poisson equations. In order to illustrate this, we consider a 1-D grid with equidistant grid spacing $\Delta$. The successive application of the operators ${\mathbb D}$ and ${\mathbb G}$ results in the stencil $\phi_i'' = (\phi_{i+2}+\phi_{i-2}-2\phi_i)/(4\Delta^2)$. Applied to a function $\phi$ whose value is zero at even-indexed positions and one at odd-indexed ones, this gives strictly zero.  

Instead, one can derive a FV discretization stencil starting from
\begin{eqnarray}
\int_{V}\nabla^2 \phi \,\mathrm{d}V  & = &  \oint_{\partial V} {\nabla \phi}\cdot \,{\mathrm{d}}{\boldsymbol S} ,\\ & = & \sum_{j \in \pi_i} \int_{S_{ij}}\nabla  \phi  \cdot \mathrm{d}{\boldsymbol S}.  \label{eq:lapl_contour}
\end{eqnarray}
\begin{figure}                 
  \begin{center}
  \setlength{\epsfysize}{7.0cm}
    \begin{tabular}{ccc}
      \setlength{\epsfysize}{7.0cm}
      \subfigure[]{\epsfbox{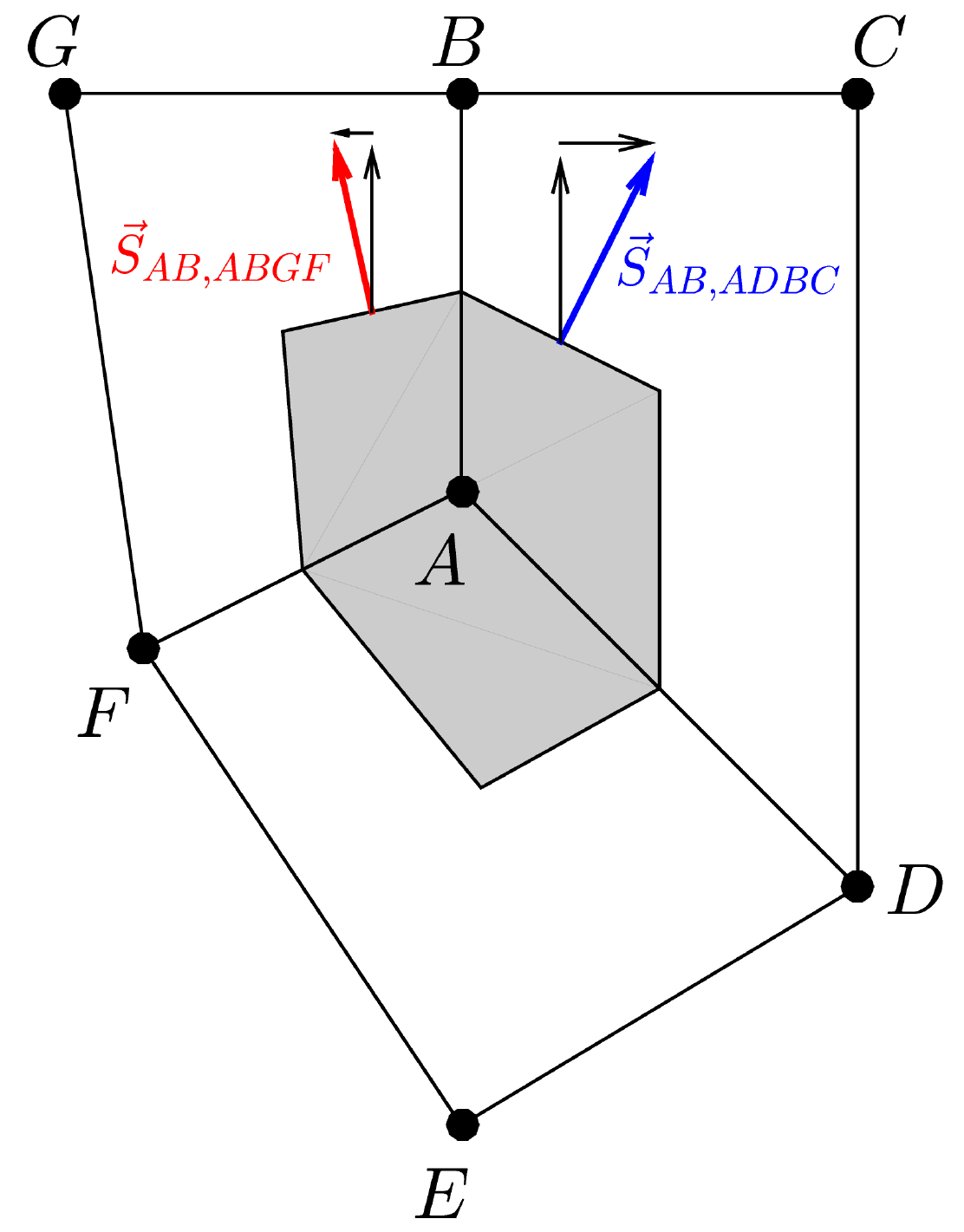}}  &
      \setlength{\epsfysize}{5.5cm}
      \subfigure[]{\epsfbox{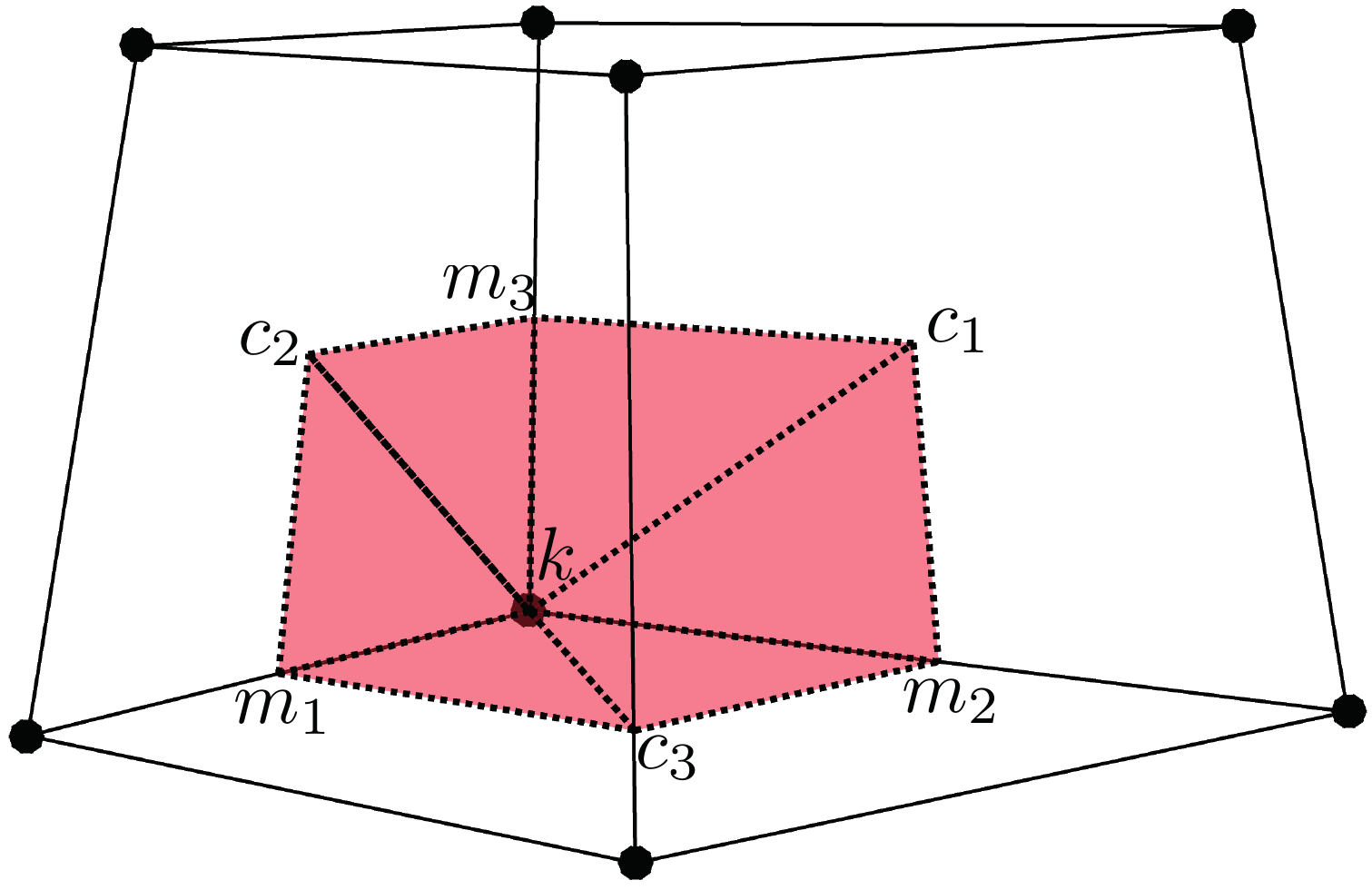}}
    \end{tabular}
    \caption{(a) Illustration of the decomposition of surface normal vectors ${\boldsymbol S}_{i,j}$ into a component parallel and perpendicular to the node pair, in case $A-B$. (b) The (red shaded) union of triangular surface patches is the elements surface $S_{fa}$ (used in expression (\ref{eq:lapl_fa})) associated with node `A'. Each of the triangular patches has the grid node $k$, a midpoint $m_i$ and a face centre $c_i$ as vertices.}
         \label{fig:lapl}                
  \end{center}
\end{figure}
 In the case of unstructured meshes, however, the interface normal ${\boldsymbol S}_{ij}$ is in general not parallel to the vector ${\boldsymbol r}_j - {\boldsymbol r}_i$. Therefore, we cannot use of the approximation $\nabla \phi = (\phi_j-\phi_i)({\boldsymbol r}_j-{\boldsymbol r}_i)/({\boldsymbol r}_j-{\boldsymbol r}_i)^2 $, as this only gives the component of $\nabla \phi$ along the direction of the vector ${\boldsymbol r}_j-{\boldsymbol r}_i$. Instead, we use a modified scheme as devised by \citet{diskin2010}. In order to introduce it, we recall that the surface associated with the node pair $i,j$ is the union of different triangular patches. This is illustrated for a 2-D example (where the surface patches are line elements) in Figure \ref{fig:lapl}; the vector ${\boldsymbol S}_{AB}$ is the sum of ${\boldsymbol S}_{AB,ADCB}$ and ${\boldsymbol S}_{AB,ABGF}$, vectors that are associated with the quadrilateral elements $ADCB$ and $ABGF$, respectively. Here, we have introduced the notation ${\boldsymbol S}_{ij,el}$ for the contribution to ${\boldsymbol S}_{ij}$ associated with triangular patches that are located within element $el$. It follows that $\sum_{el:\{i,j\} \in el} {\boldsymbol S}_{ij,el}$, and that we can recast expression (\ref{eq:lapl_contour}) as
\begin{equation}
\int_{V}\nabla^2 \phi \,\mathrm{d}V  = \sum_{j \in \pi_i}  \sum_{el:\{i,j\} \in el}  \int_{S_{ij,el}} \nabla \phi \cdot \mathrm{d}{\boldsymbol S}. \label{eq:lapl_double_sum}
\end{equation}
We can further expand the integral on the right-hand side of equation (\ref{eq:lapl_double_sum}) as 
\begin{equation}
\int_{S_{ij,el}} \nabla \phi \cdot \mathrm{d}{\boldsymbol S} = \int_{S_{ij,el}} (\nabla \phi)^{||} \cdot \mathrm{d}{\boldsymbol S}^{||} + \int_{S_{ij,el}} (\nabla \phi) \cdot \mathrm{d}{\boldsymbol S}^{\perp}. \label{eq:lapl_split}
\end{equation}
Here, $(\nabla \phi)^{||}$ and  ${\boldsymbol S}^{||}$ refer to the projection of $\nabla \phi$ and ${\boldsymbol S}$ along the edge connecting nodes $i$ and $j$, and $(\nabla \phi)^{\perp}$ to the orthogonal complement of $(\nabla \phi)^{||}$, respectively. The numerical approximation for $(\nabla \phi)^{||}$ is easily found,
\begin{equation}
(\nabla \phi)^{||}_{ij}\approx \frac{\phi({\boldsymbol r}_j) - \phi({\boldsymbol r}_i)}{\left|{\boldsymbol r}_j - {\boldsymbol r}_i \right|^2} \left({\boldsymbol r}_j - {\boldsymbol r}_i \right).
\end{equation}
For the second term in expression (\ref{eq:lapl_split}), we compute $\nabla \phi$ using a Green-Gauss approximation applied to the element $el$, i.e. we approximate:
\begin{equation}
\nabla \phi \approx \frac{1}{V_{el}}\oint_{\Omega_{el}} \phi \,\mathrm{d}{\boldsymbol S} = \sum_{fa \in el} \int_{S_{fa}} \phi \,\mathrm{d}{\boldsymbol S}. \label{eq:lapl_fa}
\end{equation}
Here ${\Omega_{el}}$ denotes the bounding surface of a grid element (like e.g. ADCB in Figure \ref{fig:lapl} (a)) and $V_{el}$ its volume. The index $fa \in el$  refers to the faces of the element $el$ and $S_{fa}$ are their respective surfaces. In order to discretize the integrals on the right-hand side of eq. (\ref{eq:lapl_fa}), we triangulate the element faces, that is, we consider them as a union of triangular surfaces that each have a grid node $k$, an edge midpoint $m$ and an element face centroid $c$ as their vertices. The surface integral of $\phi$ over $S_{fa}$ can then be expanded as a sum of surface integrals over these triangles, and each of these integrals is then approximated as:
\begin{equation}
\int _{S_{tri}} \phi \,\mathrm{d}{\boldsymbol S} \approx \frac{1}{3}\left(\Phi_k +  \Phi_m + \Phi_c \right) {\boldsymbol S}_{tri},
\end{equation}
where $S_{tri}$ denotes the triangular surface and $\Phi_{k}, \Phi_{m}$ amd $\Phi_{c}$ are the values of $\phi$ at the node $k$, edge midpoint $m$ and element face centroid $c$, respectively. These last two values are approximated by means of linear interpolation between the nodes of the face $fa$. 
All this allows us to write down a discretization of $\nabla^2 \phi$ around a grid node $i$ solely in terms of the values of $\phi$ at $i$ and its neighbours, that is,. we may write:
\begin{equation}
\left. \nabla^2 \phi \right|_{{\boldsymbol r}={\boldsymbol r_i}} \approx \left({\mathbb L} \Phi \right)_{i} = \frac{1}{V_i}\sum_{j \in \pi_i}(\mathbb{G}_{f} \Phi)_{ij} \cdot {\boldsymbol S}_{ij} = \frac{1}{V_i}\sum_{j \in \pi_i} w_{ij}\left( \Phi_{j}  - \Phi_{i} \right),
\label{eq:laplacian_final}
\end{equation}
in which the factors $w_{ij}$ are related to the grid geometry. Furthermore, we use the notation $\mathbb L$ and $(\mathbb{G}_{f} \Phi)_{ij}$ to denote the numerical approximation to the Laplacian and the face-normal component of the gradient of $\phi$ at the interface associated with the node pair $(i,j)$. 

Finally, spatial averages of a function $\phi$ over a control volume are simply approximated as
\begin{equation}
\int_{V_i} \phi \,\mathrm{d}V \approx V_{i} \Phi_{i}.
\end{equation}
This formula will be used for the discretization of a body force, like the buoyancy force we will be concerned with in section \ref{sec:benchmark}.

\subsection{Time-stepping scheme for the MHD equations \label{subsec:timestep}}
The spatial discretization procedure discussed in the previous section transforms the MHD equations into a set of ordinary differential equation for the unknown values $({\boldsymbol u}_i,{\boldsymbol b}_i,p_i)$ at the grid nodes $i$. One of the major difficulties associated with this is that the discretized magnetic induction equation does not necessarily conserve the solenoidal character of ${\boldsymbol b}$. To remedy this issue, we follow a technique pioneered by \citet{toth2000} and afterwards adopted, among others by \citet{harder2005finite}, \citet{chan2007} and \citet{guermond2007}. It consists of equipping the induction equation with a Lagrange multiplier or pseudo-pressure $p_b$ that allows projection of the magnetic field onto the space of divergence-free vector fields. More specifically, the induction equation now reads:
\begin{equation}
\frac{\partial {\boldsymbol b}}{\partial t} + {\boldsymbol u} \cdot  \nabla {\boldsymbol b} + \nabla p_b =  {\boldsymbol b}\cdot \nabla {\boldsymbol u} + Re_m^{-1} \nabla^2{\boldsymbol b}. \label{eq:induction_modif}
\end{equation} 
In order to integrate the momentum equation (\ref{eq:navierstokes}) and modified induction equation (\ref{eq:induction_modif}) in time, we adopt a fractional step algorithm akin to the one devised by \cite{chorin1968} and \cite{kim1985} for the incompressible Navier-Stokes equation as follows:
\begin{enumerate}
\item Compute the intermediate velocity and magnetic field ${\boldsymbol U}^{\star},{\boldsymbol B}^{\star}$ at the grid nodes from the given data ${\boldsymbol U}^n, {\boldsymbol B}^n, P^{n}, P^{n}_{b}$ by integration of the MHD equations, that is,
\begin{eqnarray}
\frac{{\boldsymbol U}^{\star} - {\boldsymbol U}^{n}}{\Delta t} & = & - \mathbb{C}^{n+1/2}_{U}{\boldsymbol U^{n+1/2}} + \mathbb{C}^{n+1/2}_{B}{\boldsymbol B}^{n+1/2} + \frac{1}{Re}\mathbb{L}{\boldsymbol U}^{n+1/2} -\mathbb{G}P^{n} + {\boldsymbol F}_b, \\
\frac{{\boldsymbol B}^{\star} - {\boldsymbol B}^{n}}{\Delta t} & = & - \mathbb{C}^{n+1/2}_{U}{\boldsymbol B}^{n+1/2} + \mathbb{C}^{n+1/2}_{B}{\boldsymbol U}^{n+1/2} + \frac{1}{Re_m} \mathbb{L}{\boldsymbol B}^{n+1/2} -\mathbb{G}P^{n}_{b}.
\end{eqnarray}
Here, a second order accurate Crank-Nicolson time discretisation scheme is used for the node-based, vectorial quantities, that is,
\begin{equation}
{\boldsymbol U}^{n+1/2} = \frac{{\boldsymbol U}^{\star}+{\boldsymbol U}^{n}}{2},
\end{equation}
and likewise for ${\boldsymbol B}^{n+1/2}$. For the face-centred quantities $U_{ij}$ and $B_{ij}$ that appear in the expression of the operators $\mathbb{C}_{U}$ and $\mathbb{C}_{B}$, we use an explicit second-order time accurate Adams-Bashforth scheme, that is,
\begin{equation}
U_{ij}^{n+1/2} = \frac{3}{2}U_{ij}^{n} - \frac{1}{2}U_{ij}^{n-1}, 
\end{equation}
and likewise for $B_{ij}^{n+1/2}$. This scheme has the advantage of being stable independently of the time step $\Delta t$ -- we corroborate this stability property in Appendix \ref{app:stab} -- but still being linear in the unknowns ${\boldsymbol U}^{\star}$ and ${\boldsymbol B}^{\star}$. As such, we avoid the difficulties associated with solving a nonlinear system in these quantities. The linear system is solved using a classical Jacobi iterative method. This method is preferred above a Gauss-Seidel scheme that lacks natural data parallelism in an unstructured grid setting; this is related to the fact the Gauss-Seidel method requires the inversion of an upper triangular matrix, which is essentially a sequential process. We combine the Jacobi method with a successive overrelaxation technique; in a recent work \citep{mittal}, it was shown that this technique holds the promise of a considerable speed-up of the convergence.
\item The `new' velocity ${\boldsymbol U}^{n+1}$ and magnetic field ${\boldsymbol B}^{n+1}$ are related to their intermediate counterparts  ${\boldsymbol U}^{\star}$ and ${\boldsymbol B}^{\star}$ through the following relationships:
\begin{equation}
\frac{{\boldsymbol U}^{n+1} - {\boldsymbol U}^{\star}}{\Delta t} = - \mathbb{G}\left(P^{n+1} -P^{n}\right) =  -\mathbb{G}\delta P, 
\end{equation}
\begin{equation}
\frac{{\boldsymbol B}^{n+1} - {\boldsymbol B}^{\star}}{\Delta t} = - \mathbb{G}\left(P^{n+1}_b-P^{n}_b\right) = -\mathbb{G}\delta P_b,  \label{eq:corr_magnetic}
\end{equation}
where $P$ and $P_b$ denote the discrete version of the pressure $p$ and magnetic pseudo-pressure $p_b$, respectively, and where we have introduced the shorthand notations $\delta P = P^{n+1}-P^{n}$ and $\delta P_b = P^{n+1}_b-P^{n}_b$.
Taking the divergence of this expression and imposing the incompressibility constraint on ${\boldsymbol  U}^{n+1}$ and ${\boldsymbol  B}^{n+1}$ would lead to a Poisson equation that suffers from the odd-even-decoupling described above. Instead, we enforce the solenoidality constraints at the level of the face-normal velocity and magnetic fluxes $U_{ij}$ and $B_{ij}$. We start from the intermediate face-normal velocity and magnetic field $U^{\star}_{ij}$ and $B^{\star}_{ij}$:
\begin{equation}
U^{\star}_{ij} = \frac{{\boldsymbol U}^{\star}_{i} + {\boldsymbol U}^{\star}_{j}}{2} \cdot {\boldsymbol S}_{i,j}, \hspace{1cm} B^{\star}_{ij} = \frac{{\boldsymbol B}^{\star}_{i} + {\boldsymbol B}^{\star}_{j}}{2} \cdot {\boldsymbol S}_{i,j}
\label{ustar-face}
\end{equation}
The gradients of the Lagrangian multipliers $P$ and $P_b$ at the faces can now be used to relate the new face-normal velocity $U^{n+1}_{ij}$ to $U^{\star}_{ij}$:
\begin{equation}
\frac{U^{n+1}_{ij} - U^{\star}_{ij}}{\Delta t} = \left(\mathbb{G}_{f} \delta P \right)_{ij},
\label{press-corr}
\end{equation} 
and likewise for   
\begin{equation}
\frac{B^{n+1}_{ij} - B^{\star}_{ij}}{\Delta t} = \left(\mathbb{G}_{f}\delta P_b \right)_{ij}
\label{press-corrb}
\end{equation}   
We now want to impose mass conservation at the level of the convecting velocities, that is, we want $U^{n+1}_{ij}$ and $B^{n+1}_{ij}$ to satisfy:
\begin{equation}
\sum_{j \in \pi_i } U^{n+1}_{ij} = 0
\label{incompress-np1}
\end{equation}
\begin{equation}
\sum_{j \in \pi_i } B^{n+1}_{ij} = 0
\label{incompress-np2}
\end{equation}
Combination of all this now leads to a Poisson equation for $\delta P$ and $\delta P_b$ at the CV nodes based on the discretization scheme (\ref{eq:laplacian_final}) that is free of odd-even-decoupling, and can be expressed as
\begin{equation}
\mathbb{L}\delta P = \frac{1}{\Delta t}\mathbb{D}({\boldsymbol U}^{\star}), \end{equation}
\begin{equation}
\mathbb{L}\delta P_b = \frac{1}{\Delta t}\mathbb{D}({\boldsymbol B}^{\star}). \end{equation}
These equation are solved using a BiCGstab(2)-algorithm \citep{vanderVorst} or an algebraic multigrid method \citep{hypre,boomeramg}.
\item Finally, both the nodal and face-normal velocity and magnetic field are corrected with the gradient of $\delta P$ and $\delta P_b$,
\begin{eqnarray}
{\boldsymbol U}^{n+1} &=& {\boldsymbol U}^{\star} - \Delta t \mathbb{G}\delta P, \\
U^{n+1}_{ij} & = & U^{\star}_{ij} - \Delta t \left(\mathbb{G}_f \delta P\right)_{ij}, \\
{\boldsymbol B}^{n+1} &=& {\boldsymbol B}^{\star} - \Delta t \mathbb{G}\delta P_b, \label{eq:correct_b} \\
B^{n+1}_{ij} & = & B^{\star}_{ij} - \Delta t \left(\mathbb{G}_f \delta P_b \right)_{ij}.
\end{eqnarray}
 \end{enumerate}
We note that the updating scheme for the face-centred and nodal quantities is slightly different. As a consequence of this, the nodal divergence of the velocity field is not exactly zero, but of order-of-magnitude $\mathcal{O}(\Delta x)$. Since, however, at every time step, the new nodal and face-centreed quantities are coupled by eq. (\ref{ustar-face}), the divergence of the nodal velocity can never become larger than $\mathcal{O}(\Delta x)$. As shown in Appendix \ref{app:stab}, the mixed Adams-Bashforth/Crank-Nicolson scheme ensures that the nonlinear terms exactly conserve the total energy, that is, the sum of kinetic and magnetic energy. It can be shown, on the other hand, that there is a small spurious kinetic energy dissipation term associated with the Lagrange multiplier gradients $\nabla p$ and $\nabla p_b$ that scales as $\mathcal{O}(\Delta t \Delta x^2)$ \citep{ham2007}. Therefore, the total energy cannot become unbounded, and thus we can conclude that our time-stepping method is stable.

\subsection{Validation: The incompressible Archontis dynamo}
\begin{figure}
\begin{center}               
\includegraphics[width=\columnwidth]{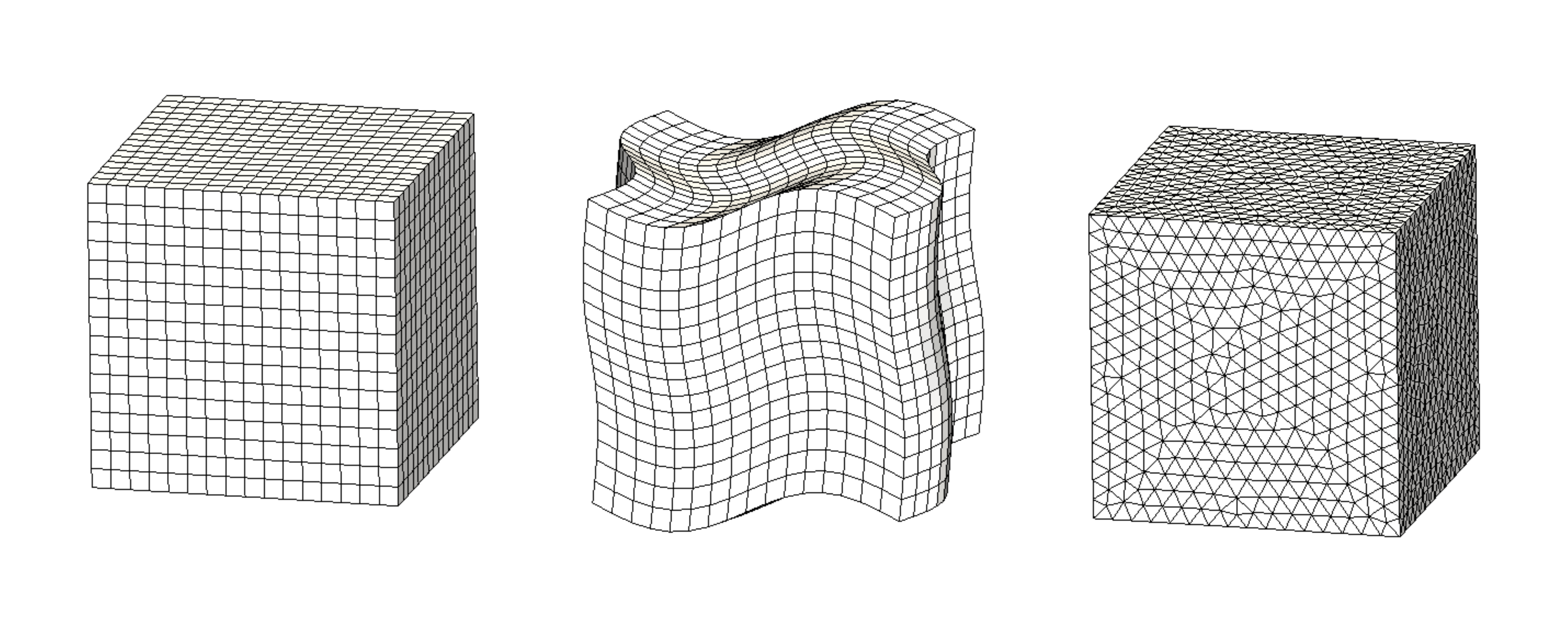}
    \caption{Illustration of the three different types of grids used for the simulation of the Archontis dynamo. In Table \ref{tab:CameronGalloway}, these are referred to as Cartesian, hexahedral and tetrahedral, respectively.}              
    \label{fig:meshes_CameronGalloway}  
  \end{center}
\end{figure}

In order to test the algorithm outlined above, we perform MHD simulations in periodic box geometries. We consider the nonlinear saturation of the Archontis dynamo \citep{ArchontisPhD}, as studied by \citet{cameron2006}. In this case, the box size is $2\pi$, and we choose $Re=Re_m=100$ and the initial conditions 
\begin{equation}
\left. {\boldsymbol u} \right|_{t=t_0} = \left. {\boldsymbol b} \right|_{t=t_0} =  \sin z \hat{\boldsymbol x} +  \sin x  \hat{\boldsymbol y}  + \sin y  \hat{\boldsymbol z} .
\end{equation}
The system is forced by a body force ${\boldsymbol f}_b = Re^{-1}\left. {\boldsymbol u} \right|_{t=t_0}$. Using a spectral method, Cameron and Galloway found that the system evolves into a stationary state in which kinetic and magnetic energy are quasi-equipartioned with 
\begin{equation}
e_{k} = \frac{1}{8\pi^3} \iiint {\boldsymbol u}^2 \,\mathrm{d}V \approx 0.1781, \hspace{1cm} e_{m} =  \frac{1}{8\pi^3} \iiint {\boldsymbol b}^2 \,\mathrm{d}V  \approx 0.1765.
\end{equation}
In Table \ref{tab:CameronGalloway}, we compare these values to the ones obtained with the present FV code. We report results for three different types of meshes, illustrated in Figure \ref{fig:meshes_CameronGalloway}, and different grid resolutions. The first type of mesh is a Cartesian one with equidistant grid spacing $\Delta x = \Delta y = \Delta z$. The second grid also consists of hexahedral elements, and is obtained as follows. Starting from a Cartesian, equidistant grid, we perturb the positions of the grid nodes according to $(x,y,z) \rightarrow (x,y,z) + 0.5(\sin(z),\sin(x),\sin(y))$. The third grid, finally, consists of isotropic tetrahedral elements and has been generated using the preprocessing software ANSYS ICEM CFD. 
\begin{table}
\begin{center}
\begin{tabular}{c|cc|cc|cc}
\hline \hline 
           &       \multicolumn{2}{|c|}{Cartesian}       &       \multicolumn{2}{|c|}{Hexahedral}    &   \multicolumn{2}{|c|}{Tetrahedral}  \\ \hline
           & $e_k$ & $e_m$ & $e_k$ & $e_m$ & $e_k$ & $e_m$ \\
$16^3$ & 0.1803 & 0.1777  &  0.1792 & 0.1767 &    \multicolumn{2}{|c|}{No dynamo}                       \\
$24^3$ & 0.1796 & 0.1775  &  0.1790 & 0.1771 & 0.1794 & 0.1776\\
$32^3$ & 0.1791 & 0.1772  &  0.1788 & 0.1769 & 0.1788 & 0.1771 \\
$48^3$ & 0.1786 & 0.1768  &  0.1784 & 0.1767 & 0.1784 & 0.1767  \\
$64^3$ & 0.1783 & 0.1766  &  0.1783 & 0.1766 & 0.1783 & 0.1766    \\
\hline \hline
\end{tabular}
\end{center}
\caption{Numerical results for the kinetic and magnetic energy density $e_k$ and $e_m$ of the Archontis dynamo at $Re=Re_m=100$. The reference values are $e_k = 0.1781$ and $e_m=0.1765$.}
\label{tab:CameronGalloway}
\end{table}

\begin{figure}
  \begin{center}
  \setlength{\epsfysize}{7.0cm}
    \begin{tabular}{ccc}
      \setlength{\epsfysize}{7.0cm}
      \subfigure[]{\epsfbox{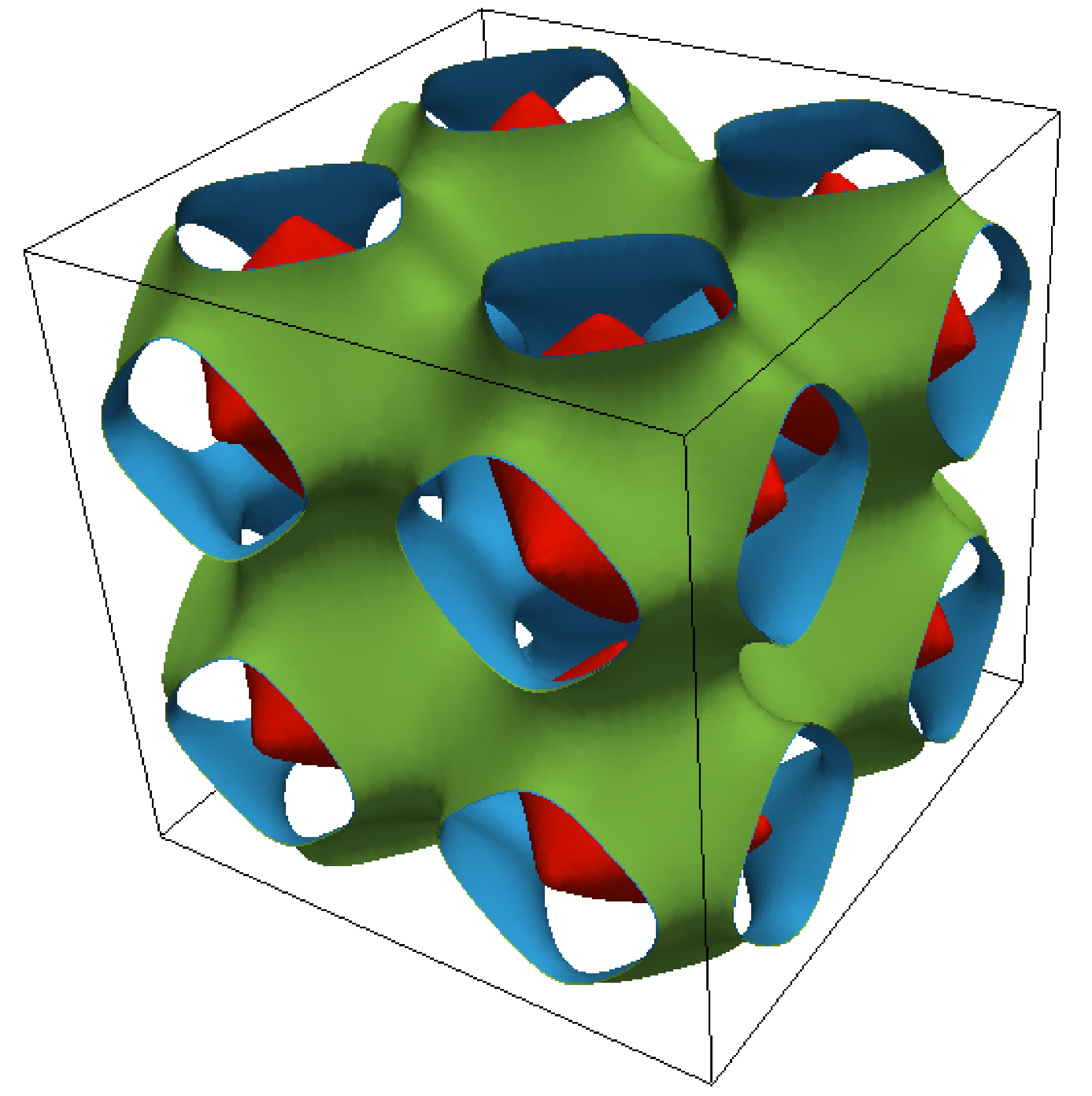}}  &
      \setlength{\epsfysize}{7.0cm}
      \subfigure[]{\epsfbox{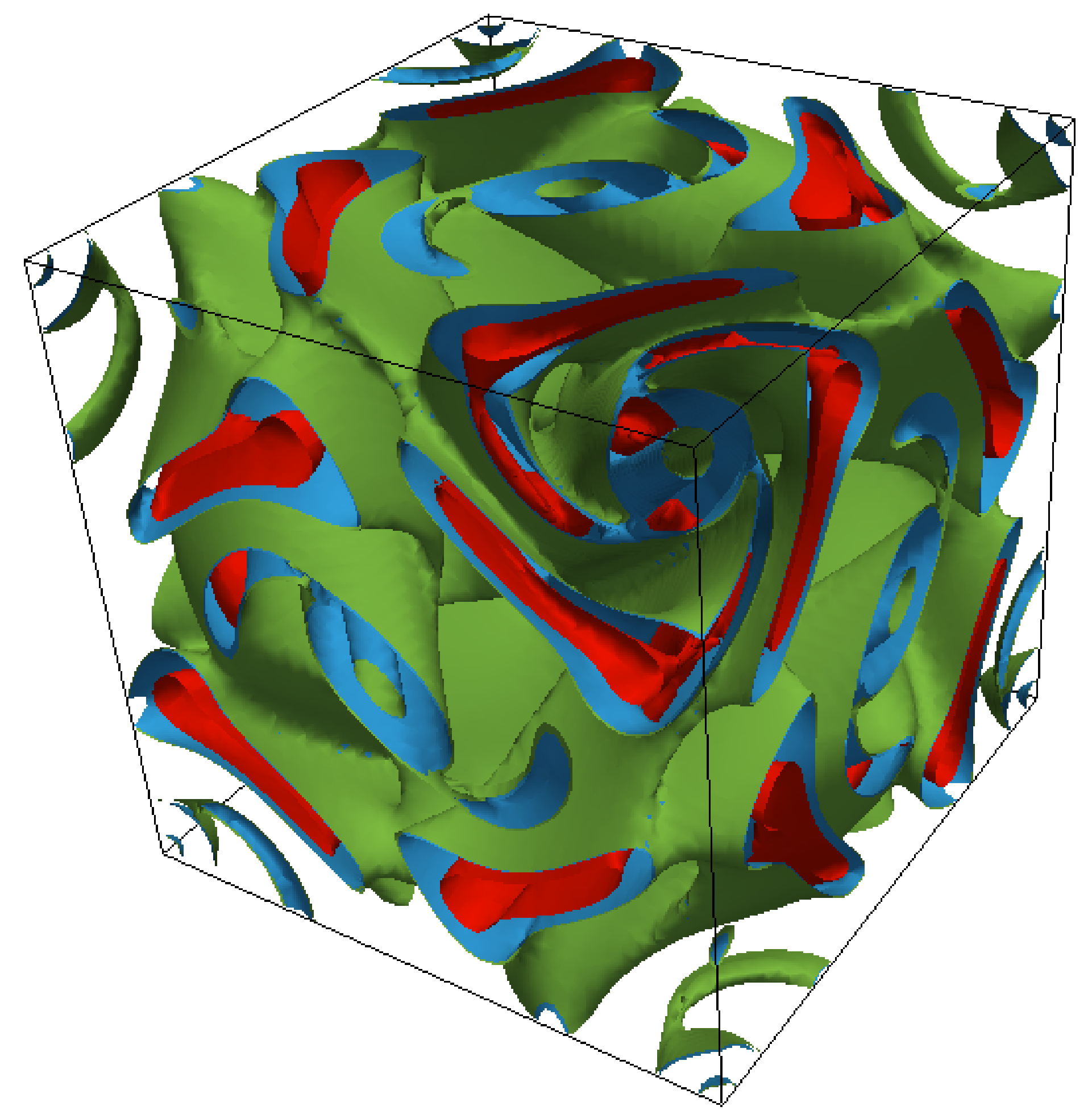}}
    \end{tabular}
    \caption{Isocontours of $({\boldsymbol u}+{\boldsymbol b})\cdot({\boldsymbol u}+{\boldsymbol b})$ (a) and $({\boldsymbol u}-{\boldsymbol b})\cdot({\boldsymbol u}-{\boldsymbol b})$ (b) for $Re=Re_m=200$. The red and blue/green isosurfaces correspond to values of 80 and 50 \% of the global maximum of the quantity, respectively. }
         \label{fig:CameronGalloway}                
  \end{center}
\end{figure}

Table 1 summarizes the results for the different grid types; the different resolutions are given in the leftmost column. We find that, by increasing the resolution, the values of $e_k$ and $e_m$ converge towards the ones reported by \citet{cameron2006}. Remarkably, the Cartesian grid does not yield results any better than the others, although this scheme is strictly second-order accurate whereas the other ones are only first order-accurate. Finally, in order to assess the fidelity of the numerical solution in a more local way, we show, in figure \ref{fig:CameronGalloway}, isocontours of the intensities of the Elsasser variables ${\boldsymbol u} + {\boldsymbol b}$ and ${\boldsymbol u} - {\boldsymbol b}$. These are essentially the same as the ones provided by \citet{cameron2006} (see their figures 4 and 5, note that we also set $Re=Re_m=200$ in agreement with the value in those figures). 
\subsection{Weak scaling tests}
\begin{figure}
\begin{center}               
\includegraphics[width=\columnwidth]{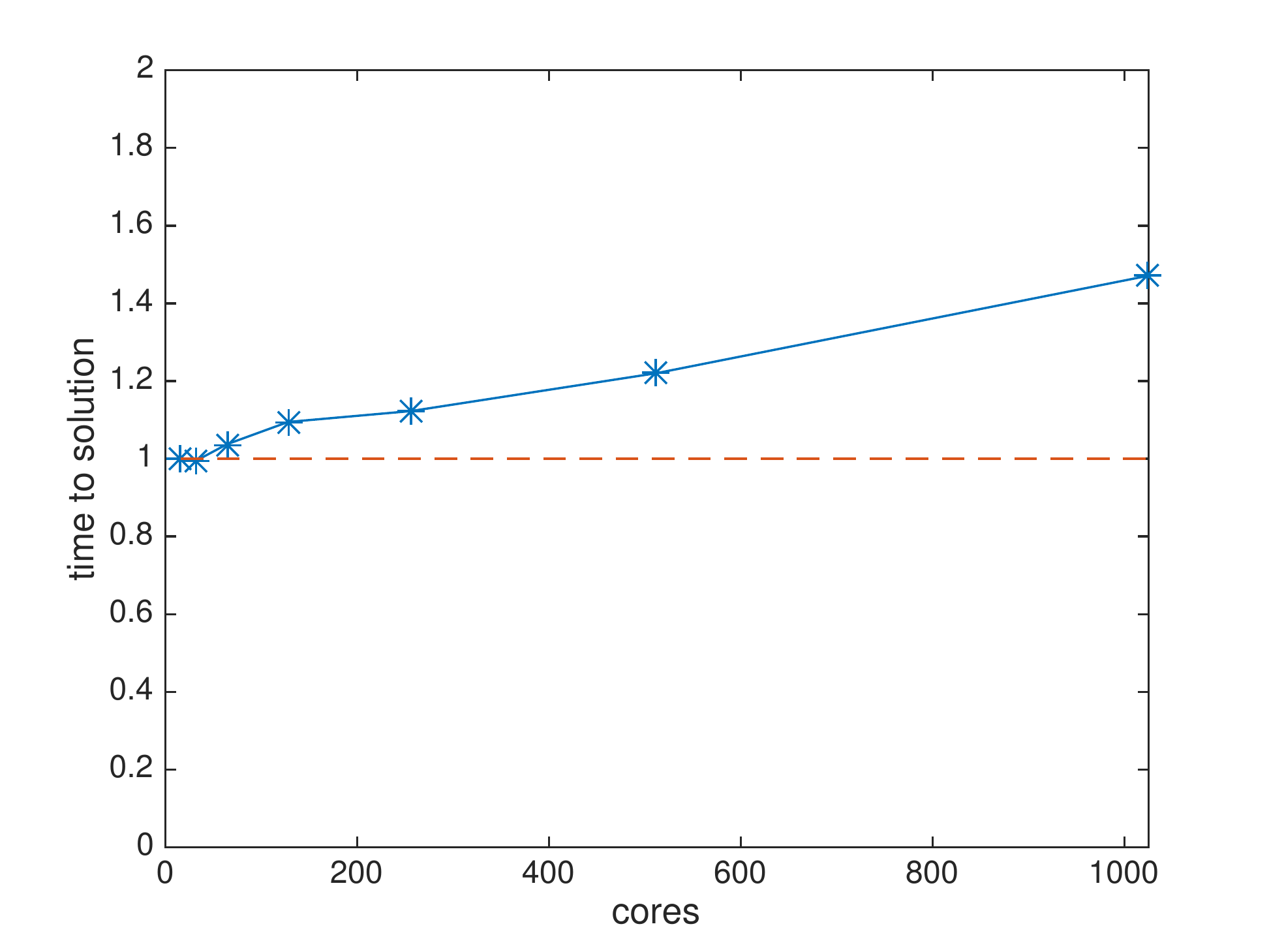}
    \caption{Weak scaling tests for a periodic box dynamo simulation. The blue symbols show the results of our scaling experiments, whereas the dashed red line indicates ideal scaling behaviour. The results have been normalized by the time-to-solution at lowest resolution (524 888 nodes, 16 cores).}              
    \label{fig:scaling}  
  \end{center}
\end{figure}

In order to investigate the computational efficiency of our code, we have carried out the so-called weak scaling tests, that is, we report the the time required to compute 500 time steps when we increase both the number of CVs and computational cores such that the number of CVs per core remains constant. We use a setup with periodic boundary conditions, similar to the ones discussed in the previous paragraph, where the number of CVs ranges from 524 888  to 33 554 432 and the number of cores from 16 to 1024. The scaling tests have been carried out on the Piz Daint system of the Swiss Supercomputing Service (CSCS). The results are shown in figure \ref{fig:scaling}. We see a slow increase in time-to-solution as we increase the number of cores, which is dominantly associated with the  higher number of iterations required to solve the Poisson equations as the number of unknowns increases.
\section{Implementation of ferromagnetic boundary conditions} \label{sec:ferro}
\subsection{Boundary conditions on the wall-normal component of ${\boldsymbol b}$}
The `bulk algorithm' having been validated, we now turn our attention to the implementation of the ferromagnetic boundary conditions ${\boldsymbol b} \times \hat {\boldsymbol n} = 0$.
As mentioned in section \ref{sec:eqns}, the main difficulty resides in the fact that our algorithm requires us to impose a condition on each of the (three) components of the magnetic field. Therefore, our implementation needs to take into account two aspects:
\begin{enumerate}
\item At all times, we enforce the tangential components of ${\boldsymbol b}$ at the boundary to be zero, and we only retain the normal component of the magnetic field.
\item We impose a boundary condition on the normal component of ${\boldsymbol b}$ that is consistent with $\nabla \cdot {\boldsymbol b} = 0$.
\end{enumerate}
We will first illustrate this for spherical geometries; for the sake of simplicity, we will assume no-slip boundary conditions for the velocity field. Adopting canonical spherical coordinates $(r,\theta,\phi)$, the divergence constraint on ${\boldsymbol b}$ reads
\begin{equation}
\frac{1}{r^2}\frac{\partial}{\partial r} \left(r^2 b_r \right) + \frac{1}{r \sin \theta} \frac{\partial}{\partial \theta}\left(b_\theta \sin \theta\right) + \frac{1}{r \sin \theta} \frac{\partial b_\phi}{\partial \phi} = 0. \label{eq:divb_spherical}
\end{equation}
The last two terms in this expression vanish since $b_\theta$ and $b_\phi$ are identically zero at the surface $r=R$. Hence, we obtain
\begin{equation}
\frac{\partial b_r}{\partial r} = -\frac{2}{R} b_r \label{eq:br_spherical}
\end{equation} 
We can use this result to implement a consistent boundary condition in the radial component of the induction equation. We start from an FV formulation of the induction equation,
\begin{equation}
\int_V \frac{\partial {\boldsymbol b}}{\partial t} \,\mathrm{d}V = \oint_{\partial V} \left\{-{\boldsymbol u}\otimes{\boldsymbol b} + {\boldsymbol b}\otimes{\boldsymbol u} + Re_m^{-1} \left( \nabla {\boldsymbol b} \right)^T  \right\} \cdot \mathrm{d}{\boldsymbol S}, 
\label{eq:induction_bnd}
\end{equation}
where the symbol $\otimes$ denotes the tensor product.

\begin{figure}
\begin{center}
\includegraphics[width=\columnwidth]{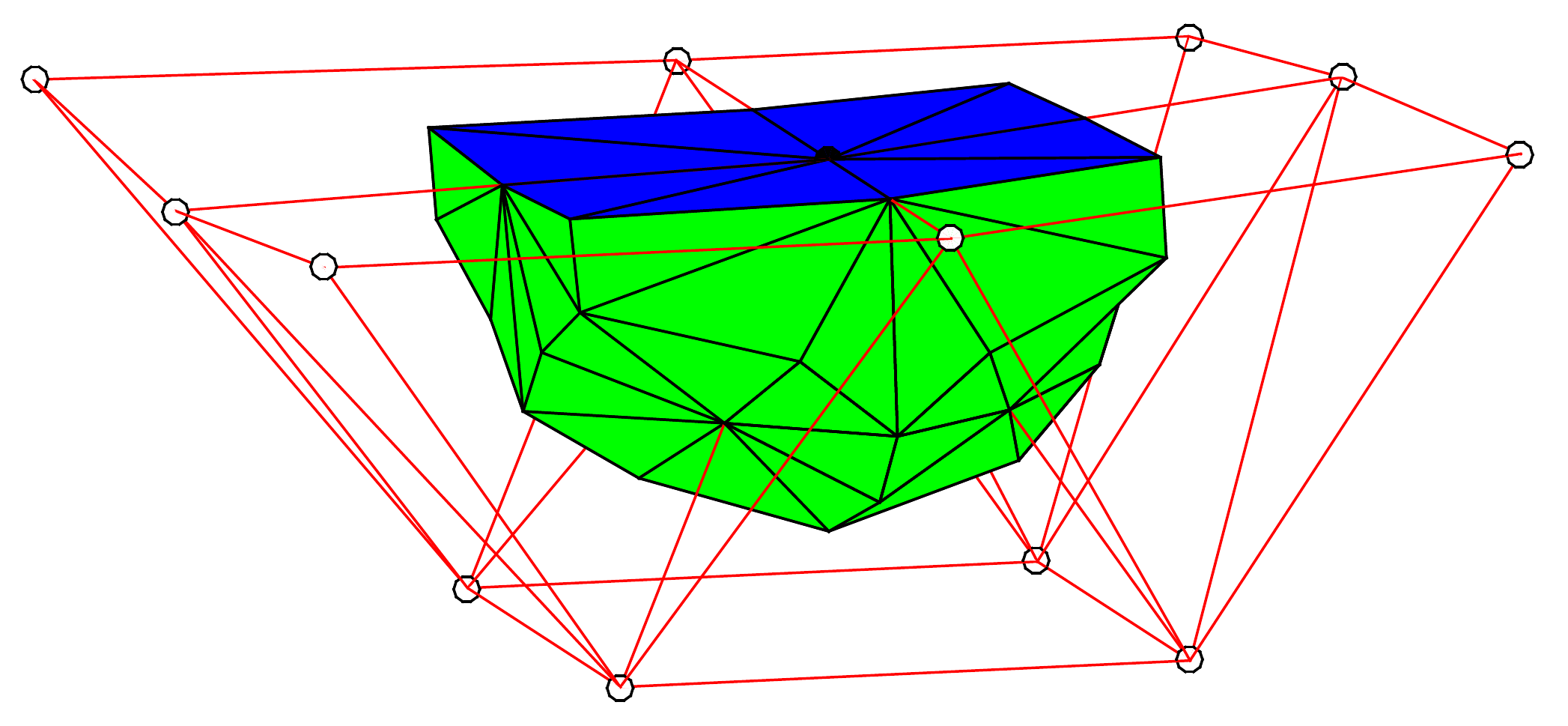}
\caption{Control volume associated with a grid node on the boundary (black-dotted). The green surface represents the internal surface patches whereas the blue one represents the boundary surface patches}
\label{fig:bndsurface}
\end{center}
\end{figure}

The integral on the right-hand side of this expression consists of a sum of surface integrals over both internal surface patches (i.e. those surface patches separating two CVs) and boundary surface patches (see Figure \ref{fig:bndsurface}). For the internal surfaces, we can use the discretization stencils laid out in Section \ref{subsec:spatial}.  The first two terms on the right-hand side of eq. (\ref{eq:induction_bnd}) do not contribute to this integral as we have assumed no-slip conditions. As we only solve for the radial component of the induction equation, it follows that we only have to compute $\hat{\boldsymbol r} \cdot \int_{S_{bnd}} (\nabla {\boldsymbol b})^{T} \cdot \mathrm{d}{\boldsymbol S}$. This quantity can be approximated as follows:
\begin{equation}
\hat{\boldsymbol r} \cdot \int_{S_{bnd}} (\nabla {\boldsymbol b})^{T} \cdot \mathrm{d}{\boldsymbol S} = \int_{S_{bnd}}(\hat{\boldsymbol r} \cdot \nabla {\boldsymbol b})^{T} \cdot \mathrm{d}{\boldsymbol S} + \mathcal{O}(\Delta x) \approx  \int_{S_{bnd}}(\hat{\boldsymbol r} \cdot \nabla {\boldsymbol b})^{T} \cdot \mathrm{d}{\boldsymbol S}.
\end{equation}
For a spherical boundary, this can be rewritten as $\int \hat{\boldsymbol r} \cdot \left( \nabla {\boldsymbol b}\right)^T \cdot \hat{\boldsymbol r} \, \mathrm{d}S$, where $\hat{\boldsymbol r}$ denotes the unit vector in radial direction. It follows that the only component of the magnetic field gradient tensor that is required to evaluate the rightmost term of eq. (\ref{eq:induction_bnd}) is $(\nabla {\boldsymbol b})_{rr} = \partial_r b_r$; thus, no tangential derivatives are required to compute the diffusive flux. Moreover, we can use eq. (\ref{eq:br_spherical}) to eliminate the (unknown) $\partial_r b_r$ and use directly the values from $b_r$ in the computation of the diffusive flux through the boundary surface patch. Summarizing, the pseudo-vacuum boundary condition gives rise to an additional diffusive boundary term of the form $-\int 2 Re_m^{-1} b_r /R\,\mathrm{d}S$ in the radial component of the induction equation. 

Motivated by the recent interest in dynamo action driven by tidal effects \citep[e.g.,][]{arkani2008tidal,arkani2009did}, we now generalize this approach towards ellipsoidal boundaries, which are described by:
\begin{equation}
\frac{x^2}{a^2} + \frac{y^2}{b^2} + \frac{z^2}{c^2} = 1. \label{eq:def_ellipse}
\end{equation}
It will be instructive at this point to introduce the (orthogonal) confocal ellipsoidal coordinate system $(\lambda,\mu,\nu)$, where each of $\lambda,\mu,\nu$ is a root $\chi$ of:
\begin{equation}
\frac{x^2}{a^2+\chi}+\frac{y^2}{b^2+\chi}+\frac{z^2}{c^2+\chi} = 1. \label{eq:def_coordinate}
\end{equation}
Without loss of generality, we can choose the isosurface $\lambda=0$ to represent the ellipsoidal surface (\ref{eq:def_ellipse}) such that the unit vector $\hat{\boldsymbol \lambda}$ points in wall-normal direction and the ferromagnetic boundary condition is equivalent to $b_{\mu} = b_{\nu} = 0$. An argument  similar to the one invoked for  the spherical case shows that we can restrict ourselves to one single element of the magnetic field gradient tensor to compute the diffusive flux across the boundary surface, more specifically $\left( \nabla {\boldsymbol b} \right)_{\lambda \lambda}$.
Using standard tensor calculus in orthogonal curvilinear coordinates, together with $b_{\mu} = b_{\nu} = 0$, we find that this component is given by
\begin{equation}
\left( \nabla {\boldsymbol b} \right)_{\lambda \lambda} = \frac{\partial b^{\lambda}}{\partial \lambda} + \Gamma^{\lambda}_{\lambda \lambda} b^{\lambda}, \label{eq:hessian_ll}
\end{equation}
where $\Gamma$ denotes a Christoffel symbol of the second kind and superscripts refer to contravariant vector components. As in the spherical case, we see that eq. (\ref{eq:hessian_ll}) does not contain any tangential derivatives. The solenoidal constraint  on the other hand reads (in contravariant form)
\begin{equation}
\frac{\partial b^{\lambda}}{\partial \lambda} + \left(\Gamma^{\lambda}_{\lambda \lambda} + \Gamma^{\mu}_{\lambda \mu} + \Gamma^{\nu}_{\lambda \nu} \right) b^{\lambda}=0.
\end{equation}
This allows the recasting of eq. (\ref{eq:hessian_ll}) as 
\begin{equation}
\left( \nabla {\boldsymbol b} \right)_{\lambda \lambda} = - \left( \Gamma^{\mu}_{\lambda \mu} + \Gamma^{\nu}_{\lambda \nu} \right) b^{\lambda}. \label{eq:christoffel}
\end{equation}
We are now left with finding an expression for the Christoffel symbols $\Gamma^{\mu}_{\lambda \mu}$ and $ \Gamma^{\nu}_{\lambda \nu} $. We can invoke the well-known identity   
\begin{equation}
\Gamma^{\mu}_{\lambda \mu} = \frac{\partial}{\partial \lambda} \log h_{\mu}, \Gamma^{\nu}_{\lambda \nu} = \frac{\partial}{\partial \lambda} \log h_{\nu}
\end{equation}
The scale factors $h_{\lambda}, h_{\mu}$ and $h_{\nu}$ \citep{dassios} are
\begin{equation}
h_{\lambda} = \frac{1}{2}\sqrt{\frac{(\lambda-\nu)(\lambda-\mu)}{(a^2+\lambda)(b^2+\lambda)(c^2+\lambda)}},
\end{equation}
\begin{equation}
h_{\mu} = \frac{1}{2}\sqrt{\frac{(\mu-\nu)(\mu-\lambda)}{(a^2+\mu)(b^2+\mu)(c^2+\mu)}},
\end{equation}
\begin{equation}
h_{\nu} =  \frac{1}{2}\sqrt{\frac{(\nu-\mu)(\nu-\lambda)}{(a^2+\nu)(b^2+\nu)(c^2+\nu)}} .
\end{equation}
After some algebra, we eventually find
\begin{equation}
\Gamma^{\mu}_{\lambda \mu} + \Gamma^{\nu}_{\lambda \nu} = -\frac{1}{2} \frac{\mu + \nu}{\mu \nu}.
\end{equation}
In order to recast the numerator and denominator of the above expression, we start from expression (\ref{eq:def_coordinate}) which defines a cubic equation in $\chi$, and thus can be written as
\begin{equation}
\chi^3 + \alpha_2 \chi^2 + \alpha_1 \chi + \alpha_0 = 0.
\end{equation}
On the ellipsoidal boundary surface, the term $\alpha_0=0$ as we know that $\chi_1 = \lambda = 0$. Thus, we have
\begin{equation}
\chi\left(\chi^2 + \alpha_2 \chi + \alpha_1 \right) = 0.
\end{equation}
The sum and product of the two roots, $\chi_{2,3}=\left\{ \mu,\nu \right\}$, of the quadratic polynomial between brackets are $-\alpha_2$ and $\alpha_1$, respectively. After some algebra, we obtain
\begin{equation}
\Gamma^{\mu}_{\lambda \mu} + \Gamma^{\nu}_{\lambda \nu} = \frac{1}{2}\frac{\alpha_2}{\alpha_1} = \frac{1}{2} \frac{a^2+b^2+c^2-x^2-y^2-z^2}{a^2b^2+a^2c^2+b^2c^2-(b^2+c^2)x^2-(a^2+c^2)y^2-(a^2+b^2)z^2}. \label{eq:christoffel2}
\end{equation} 
Finally, we note that that the contravariant component ${b^{\lambda}}=h_\lambda^{-1} {\boldsymbol b} \cdot \hat {\boldsymbol \lambda}$. This, together with expressions (\ref{eq:christoffel}) and (\ref{eq:christoffel2}), gives us all elements required for the computation of the diffusive flux through an ellipsoidal surface in terms of ${\boldsymbol b} \cdot \hat {\boldsymbol \lambda}$. 
In the specific case of a spherical surface with $a^2=b^2=c^2=x^2+y^2+z^2=R^2$, we find 
\begin{equation}
\hat{\boldsymbol \lambda} \cdot \left\{\int (\nabla {\boldsymbol b})^T\cdot \mathrm{d}{\boldsymbol S}\right\} \approx \int \hat{\boldsymbol \lambda} \cdot  (\nabla {\boldsymbol b})^T\cdot \mathrm{d}{\boldsymbol S} \approx -\frac{1}{R^2}\int b^{\lambda} \mathrm{d}S = -\frac{1}{R^2}\int h_\lambda^{-1} {\boldsymbol b} \cdot \hat{\boldsymbol \lambda} \mathrm{d}S = -\frac{2}{R} \int {\boldsymbol b} \cdot \hat{\boldsymbol r} \mathrm{d}S,
\end{equation}
which is consistent with the result previously derived in terms of spherical coordinates.

We now have obtained consistent boundary conditions for the wall-normal component of the (physical) magnetic field. The time-stepping approach outlined in Section \ref{subsec:timestep}, however, requires us to define boundary conditions on the `intermediate' magnetic field ${\boldsymbol b}^{\star}$ and the magnetic pseudo-pressure $p_b$. We choose to apply the same condition on ${\boldsymbol b}^{\star}$ as ${\boldsymbol b}^{n+1}$, that is ${\boldsymbol b}^{\star} \times \hat{\boldsymbol n} = {\boldsymbol b}^{\star} \times \hat{\boldsymbol n}=0$. This, combined with (\ref{eq:corr_magnetic}), implies that the correct boundary condition on $p_b$ is 
$\nabla p_b \times \hat{\boldsymbol n}=0$, that is, a Dirichlet condition on $p_b$.

\subsection{Magnetic decay modes}
\begin{figure}
\begin{center}\includegraphics[width=\columnwidth]{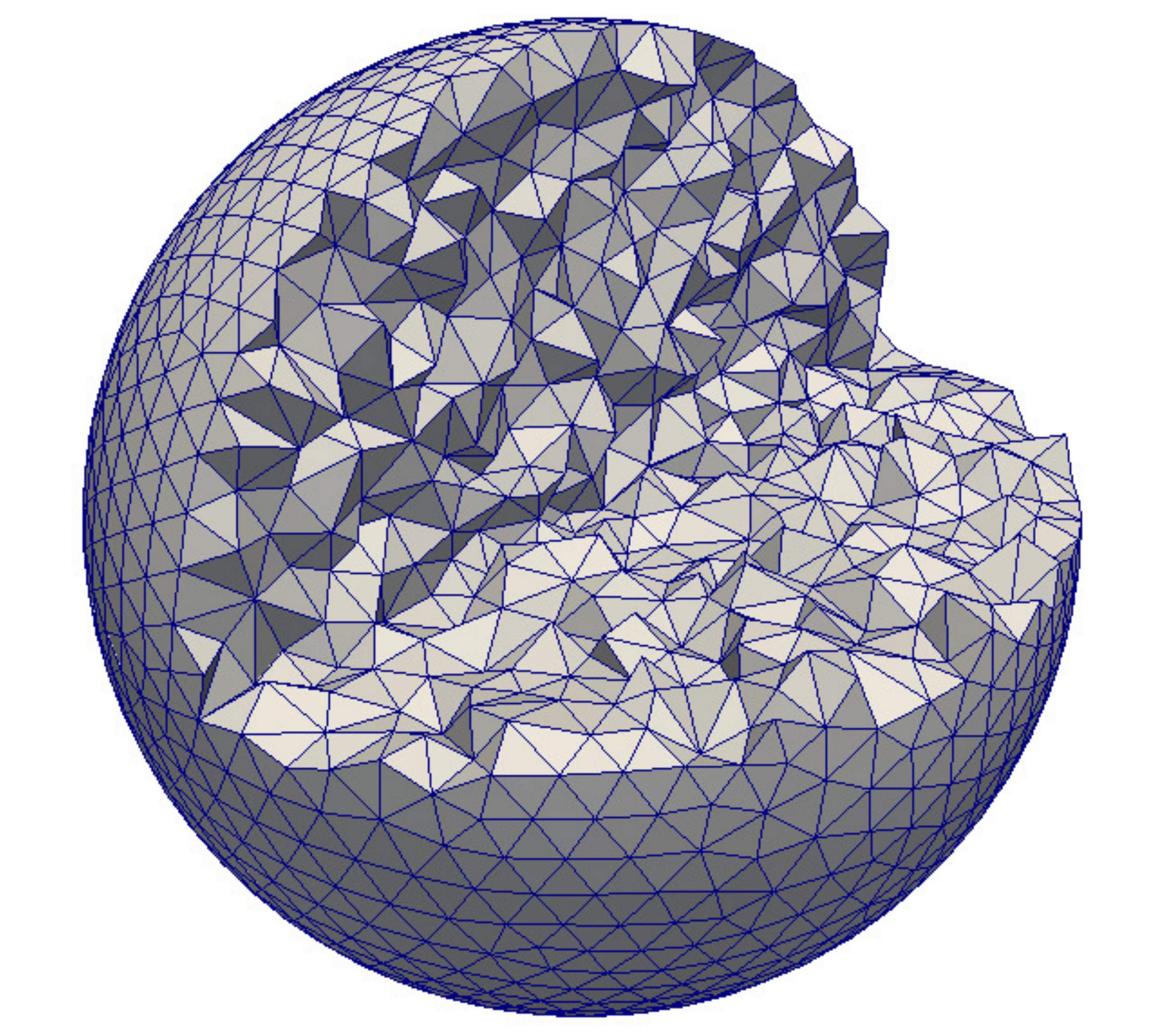}
    \caption{Illustration of the tetrahedral grid system used for the simulation of the magnetic diffusion equation in the full sphere/spheroid/ellipsoid geometry.}              
    \label{fig:gridtetra}  
  \end{center}
\end{figure}

To verify the procedure outlined above, we perform simulations of the magnetic diffusion equation, which can, without loss of generality, be written as follows:
\begin{equation}
\frac{\partial {\boldsymbol b}}{\partial t } = \nabla^2{\boldsymbol b}.
\end{equation}
This equation has an infinite set of eigensolutions ${\boldsymbol b}_i = {\boldsymbol B}_i({\boldsymbol r})\exp(-\sigma_i t)$ with the eigenvalues ${\sigma_i}$ being real and strictly positive. Analytical solutions in spherical domains can be easily found (e.g. \cite{sheyko_thesis} and Appendix \ref{app:decay}). We find $\sigma = 7.527926$ and $20.19064$ as the analytical decay rates of the slowest decaying poloidal and toroidal mode in a full sphere of radius one. We compare these results against decay rates obtained with the present FV code for two different initial conditions, which correspond to a purely poloidal and toroidal field, respectively. We use a grid that consists of tetrahedral elements (illustrated in Figure \ref{fig:gridtetra}) and a time step $\Delta t = 5 \cdot 10^{-3}$. As shown in Table \ref{tab:fullsphere}, the decay rates obtained with the FV code clearly converge towards the analytical ones as the resolution is increased.
\begin{table}
\begin{center}
\begin{tabular}{c|ccc}
\hline \hline 
     $\#$ CVs        &  Poloidal                             &    Toroidal       &  Random \\ \hline
$5844$                & 7.522178                           &  19.95080        &  7.521235                         \\
$40327$              & 7.524224                           &  20.12600        &  7.524335                      \\
$156673$            & 7.525911                           &  20.16512       &   7.525112                       \\
\hline
Analytical             & 7.527926                          & 20.19064 & 7.527926               \\
\hline \hline
\end{tabular}
\end{center}
\caption{Decay rates of a magnetic field in a full sphere with pseudo-vacuum boundary conditions associated for a purely toroidal, poloidal and random initial condition. }
\label{tab:fullsphere}
\end{table}
Further tests have been performed with a random initial magnetic field.  We first generate a random field that satisfies neither the divergence constraint nor the boundary condition. A consistent initial condition is then obtained by using a Lagrange multiplier that projects the original field onto the subspace of solenoidal fields that satisfy the boundary condition (\ref{eq:bcb}). Precise values of the decay rates for different resolutions are shown in the rightmost column of Table \ref{tab:fullsphere}. We find that these agree well with the values obtained for a purely poloidal initial condition.

Similar tests were carried out for a number of other geometries such as full spheroids and ellipsoids and also shell geometries, and the results are summarized in Table \ref{tab:decay_geometries}. Analytical results are only available for the spheical shell  (see Appendix \ref{app:decay}) and the axisymmetric toroidal mode of the full spheroid \citep{wu2009dynamo}; for the latter geometry we can rely on the fact that the boundary condition for the toroidal mode is the same for an insulating as for a ferromagnetic exterior. We note that, in a spheroidal geometry, only the axisymmetric eigenmodes are purely poloidal or toroidal.  In those cases where no analytical result is available, we have benchmarked our code against a finite-element code (D. C\'ebron, personal communication) that uses a vector potential formulation\footnote{The vector potential associated with a magnetic field ${\boldsymbol b}$ is a vector field ${\boldsymbol a}$ such that ${\boldsymbol b} = \nabla \times {\boldsymbol a}$.}. We note also that it is not possible to separate between toroidal and poloidal modes in the case of a non-axisymmetric ellipsoidal geometry. Therefore we report only one value for the slowest decay rate.  Overall, we find that there is a very good agreement between the different approaches. Given that the numerical methods are fairly different, we can assume that this validates our implementation of the ferromagnetic boundary conditions.

\begin{table}
\begin{center}
\begin{tabular}{cc|cc|cc|c}
\hline \hline 
Geometry             &      (outer) Semi-axes         &    $\sigma_{FV}^{P}$ & $\sigma_{FV}^{T}$  &    $\sigma_{ref}^{P}$    & $\sigma_{ref}^{T}$  & Reference type\\ \hline
Spherical shell    &      $(20/13,20/13,20/13)$       &    2.2207            &   10.642                       & 2.2279    & 10.634                   & Analytical         \\
Full spheroid        &      $(1,1,0.8)$    &    7.6933            &  22.376           & 7.6962    & 22.412        & FE (P), analytical (T)                    \\
Spheroidal shell  &      $(20/13,20/13,16/13)$   &    2.4303            &  13.203           & 2.4475    & 13.177 &FE                      \\
Full ellipsoid        &       $(\sqrt{1.44},\sqrt{0.56},1)$           & 9.1656 &        &   9.1728 &          &      FE                        \\
\hline \hline
\end{tabular}
\end{center}
\caption{Decay rates for the slowest decaying magnetic field eigenmodes in different geometries. The shell geometries are homothetic with the ratio between inner and outer radii $r_i/r_o=0.35$. The superscripts `P' and `T' refer to poloidal and toroidal eigenmodes, respectively. FE refers to a finite-element solution implemented in Comsol (D. C\'ebron, personal communication). For the full ellipsoid, we report the slowest decaying mode, it is neither P or T.}
\label{tab:decay_geometries}
\end{table}

\section{Self-consistent convection-driven dynamo benchmark }\label{sec:benchmark}
The final and most challenging benchmark exercise concerns a convection-driven dynamo simulation that provides a simplified model for rotating planetary cores or solar convection zones. We use a Boussinesq approximation and the geometry is a spherical shell with inner radius $r_i=7/13$ and outer radius $r_o=20/13$. The equations governing this system can then be written in the following non-dimensional form:
\begin{eqnarray}
\frac{\partial \Theta}{\partial t} + {\boldsymbol u}\cdot{\nabla} \Theta & = & q \nabla^2  \Theta, \label{eq:temp} \\
\nabla \cdot {\boldsymbol u} & = & 0, \\
Ro \left( \frac{\partial {\boldsymbol u}}{\partial t} + {\boldsymbol u}\cdot {\nabla} {\boldsymbol u} \right)+ \hat{\boldsymbol z}\times{\boldsymbol u} +\nabla p & = &  E\nabla^2{\boldsymbol u} + {\boldsymbol b}\cdot \nabla {\boldsymbol b} + q Ra \Theta {\boldsymbol r}, \\
\frac{\partial {\boldsymbol b}}{\partial t} + {\boldsymbol u}\cdot {\nabla} {\boldsymbol b} & = &  {\boldsymbol b}\cdot {\nabla} {\boldsymbol u} + \nabla^2 {\boldsymbol b}, \\
\nabla \cdot {\boldsymbol b} & = & 0. \label{eq:divb2}
\end{eqnarray}
Compared to the set of equations (\ref{eq:divu})-(\ref{eq:divb}), we note the presence of the terms $\hat{\boldsymbol z} \times {\boldsymbol u}$ and $Ra \Theta$ in the momentum equation. Note also that we use a different non-dimensionalization compared to eqs. (\ref{eq:divu})-(\ref{eq:divb}), for example, time is measured in units of magnetic diffusion time \citep{jackson2014}.
These embody the Coriolis force and a thermally driven buoyancy force, respectively. Overall, four non-dimensional parameters are required to characterize the system: the Roberts number $q$, the Rayleigh number $Ra$, the Ekman number $E$ and the Rossby number $Ro$. They are related to the physical properties of the system by the following definitions:
\begin{equation}
Ro = \frac{\eta}{2 \Omega d^2}, E=\frac{\nu}{2\Omega d^2}, Ra = \frac{g\alpha \Delta \Theta d}{2 \Omega \kappa}, q = \frac{\kappa}{\eta},
\end{equation}
where $\Omega$ denotes the spin rate, $d=r_o-r_i$ the shell thickness, $\kappa$ the heat diffusivity, $\eta=(\mu \sigma)^{-1}$ the magnetic diffusivity, $g$ gravity and $\alpha$ the thermal expansion coefficient of the fluid, which are all constant and uniform. $\Delta \Theta $ is a measure of
 the temperature contrast across the shell. To close the system, the equations need to be augmented with suitable boundary conditions. For the velocity, we prescribe a no-slip boundary condition. The magnetic field obeys the pseudo-vacuum condition (\ref{eq:bcb}), and we impose a fixed value of the temperature such that the fluid has an unstable stratification, more specifically
\begin{equation}
\Theta(r_o) = 0 < \Theta(r_i) = 1.
\end{equation}
The first solution of the system (\ref{eq:temp})-(\ref{eq:divb2}) using local methods was provided by \cite{harder2005finite} for the parameter sets $E = 5 \cdot 10^{-4}, Ra=32.5$, $q=\{4,5,8\}$ and $Ro=5\cdot 10^{4}/q$.
 They reported the occurrence of a subcritical dynamo for $q=\{5,8\}$. The case $q=5$ was later the base for a natural dynamo community benchmark exercise by \cite{jackson2014}. \citet{sheyko_thesis} studied the case $q=8$ using a pseudospectral code for a large set of initial magnetic field intensities associated with the field configuration
\begin{eqnarray}
b_r & = & \frac{1}{\sqrt{2}} \frac{5}{8} \frac{9r^3-4(4+3(r_i+r_o))r^2 + (4r_o + r_i(4+3r_o))6r - 48r_i r_o}{r}\cos \theta, \\
b_{\theta} &=& -\frac{1}{\sqrt{2}} \frac{15}{4}\frac{(r-r_i)(r-r_o)(3r-4)}{r} \sin \theta, \\
b_{\phi} &=& \frac{1}{\sqrt{2}} \frac{15}{8} \sin \pi(r-r_i) \sin 2 \theta,
\end{eqnarray}
The initial temperature distribution is
\begin{equation}
\Theta = \frac{r_o r_i}{r} - r_i + \frac{21}{\sqrt{17920 \pi}}(1 - 3\xi^2 + 3\xi^4 - \xi^6) \sin^4 \theta \cos 4\phi,
\end{equation}
where $\xi = 2r -r_i - r_o$, and the initial velocity is ${\boldsymbol u}={\boldsymbol 0}$. It is also useful at this point to define the kinetic energy $E_k$ and magnetic energy $E_m$ as follows:
\begin{equation}
E_{k} = \frac{1}{2}\iiint {\boldsymbol u}^2 \,\mathrm{d}V, 
\end{equation}
\begin{equation}
E_{m} = \frac{1}{2 Ro }\iiint {\boldsymbol b}^2 \,\mathrm{d}V. 
\end{equation}

\begin{figure}                 
  \begin{center}
        \includegraphics[width=\columnwidth]{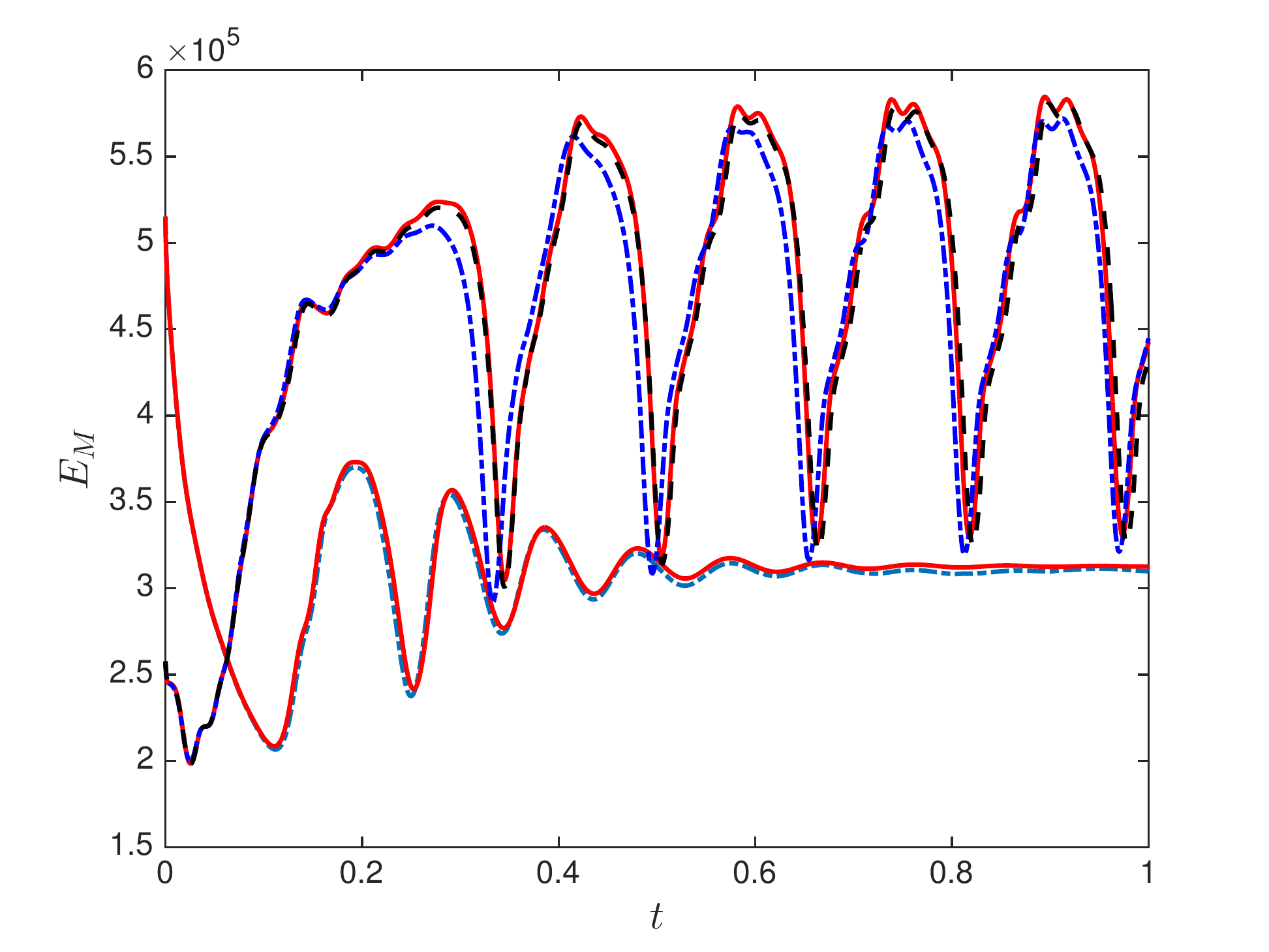}
        \caption{Time series of the magnetic energy $E_m$ for the spectral (blue, dot-dashed) and FV code with $6\times64^3$ CVs (red, solid) for two different initial magnetic field intensities. For the oscillating solution, an FV run at higher resolution ($6 \times 128^3$ CVs) is shown as well (black, dashed). Time $t$ is measured in units of magnetic diffusion time.}
         \label{fig:benchmarkPm8}                
  \end{center}
\end{figure}

As shown in Figure \ref{fig:benchmarkPm8}, it was found by \citet{sheyko_thesis} that, for $q=8$, two different types of self-sustained dynamo behaviour can occur depending on the initial conditions. For initial values of the magnetic energy between $407101$ and $623428$, a quasi-steady dynamo is found, that is, a solution that is time-independent apart from a steady azimuthal drift. In other terms, the solution can be expressed in the form $\left( {\boldsymbol u},{\boldsymbol b}, \Theta \right)=f(r,\theta,\phi-\omega t)$. Outside this range, the dynamo is not quasi-steady anymore, but exhibits relaxation oscillations similar to those observed by \cite{busse2006}.

\begin{figure}
\begin{center}               
\includegraphics[width=\columnwidth]{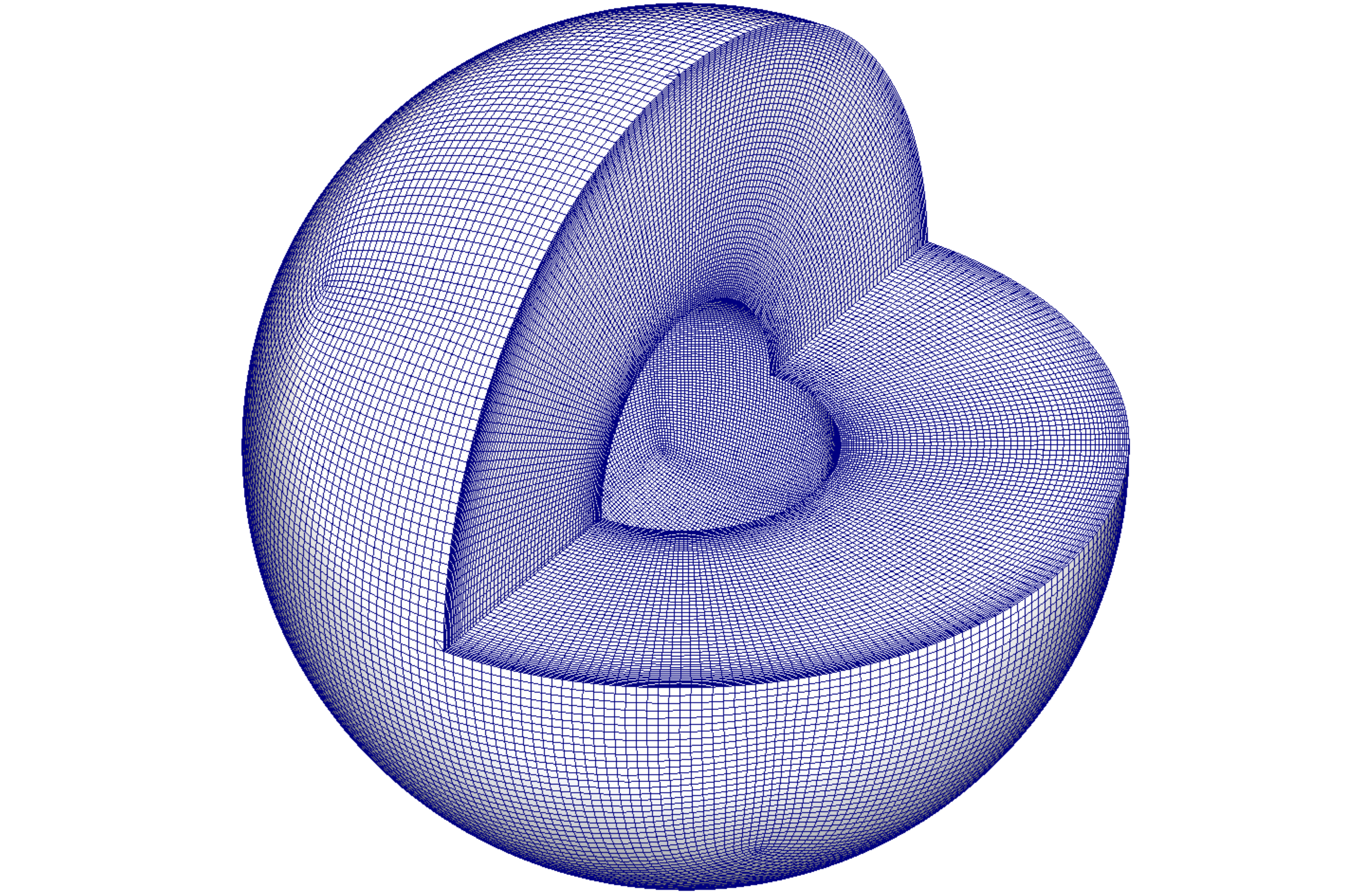}
    \caption{Illustration of the cubed grid system used for the convection-driven dynamo system. Note also the stretching of the radial grid point distribution.}              
    \label{fig:cubedsphere}  
  \end{center}
\end{figure}

In Figure \ref{fig:benchmarkPm8}, we compare results obtained with the present FV code and a reference code for the choice $q=8$. For the FV simulation, we use a so-called cubed-sphere grid system that consists of six equal blocks of hexahedral elements (see Figure \ref{fig:cubedsphere}). We consider two different resolutions, that consist of blocks of $64^3$ and $128^3$ CVs, respectively. The grid is stretched in the radial direction in order to properly resolve the Ekman boundary layers whose thickness scales as $E^{1/2}$. More specifically, the stretching is such that we have at least five CVs within a radial distance $E^{1/2}$ from the boundaries. A Crank-Nicolson scheme is used to discretise the Coriolis and buoyancy terms, and the time step $\Delta t = 2 \cdot 10^{-5}$. The reference code on the other hand is pseudospectral in angle and uses finite-differences in radius. The nonlinear and Coriolis term are integrated in time by means of a second-order accurate predictor-corrector scheme, whereas a Crank-Nicolson scheme is applied for the diffusive terms.

\begin{figure}                 
  \begin{center}
  \setlength{\epsfysize}{7.0cm}

    \begin{tabular}{ccc}
      \setlength{\epsfysize}{7.0cm}
      \subfigure[]{\epsfbox{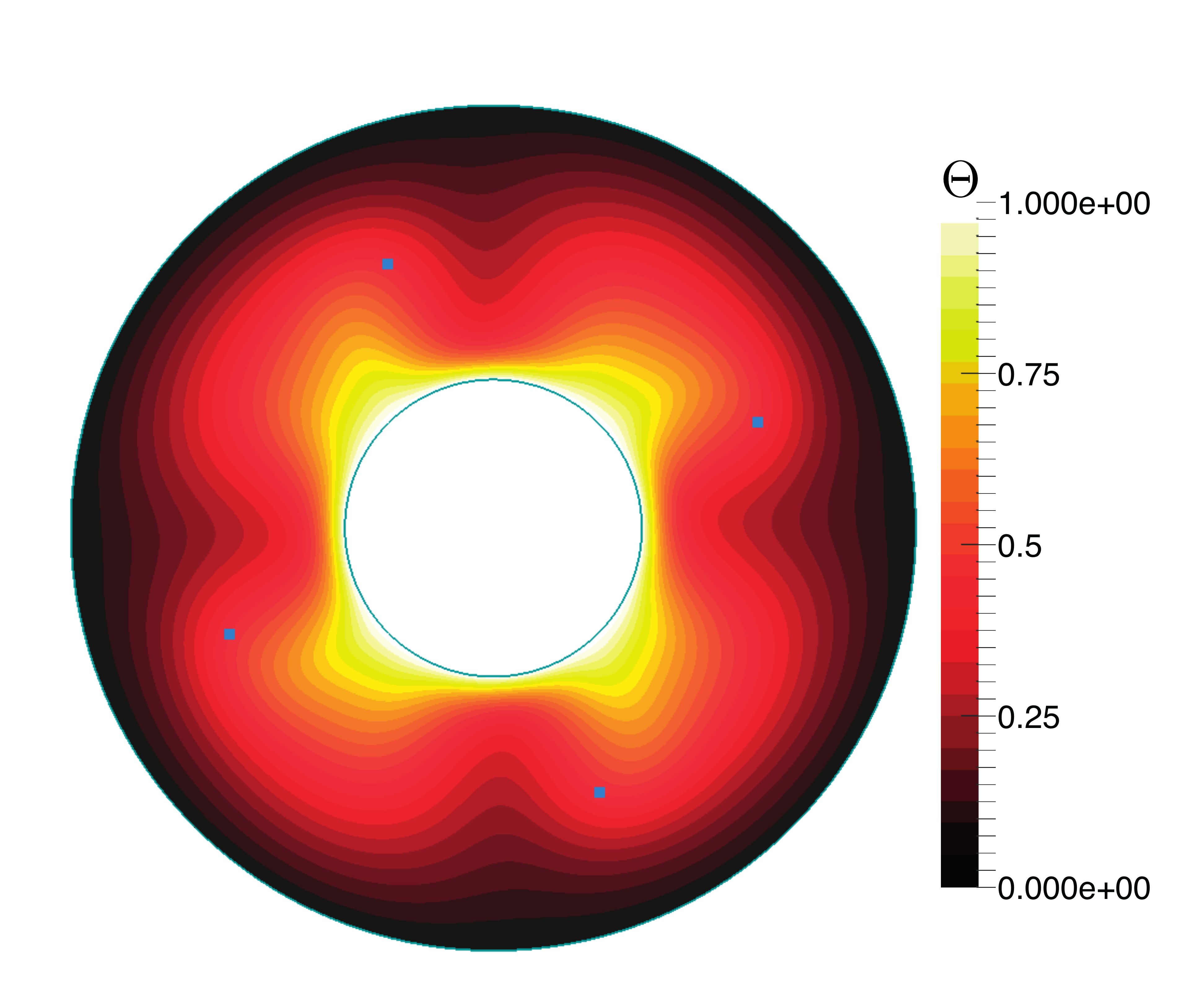}}  &
      \setlength{\epsfysize}{7.0cm}
      \subfigure[]{\epsfbox{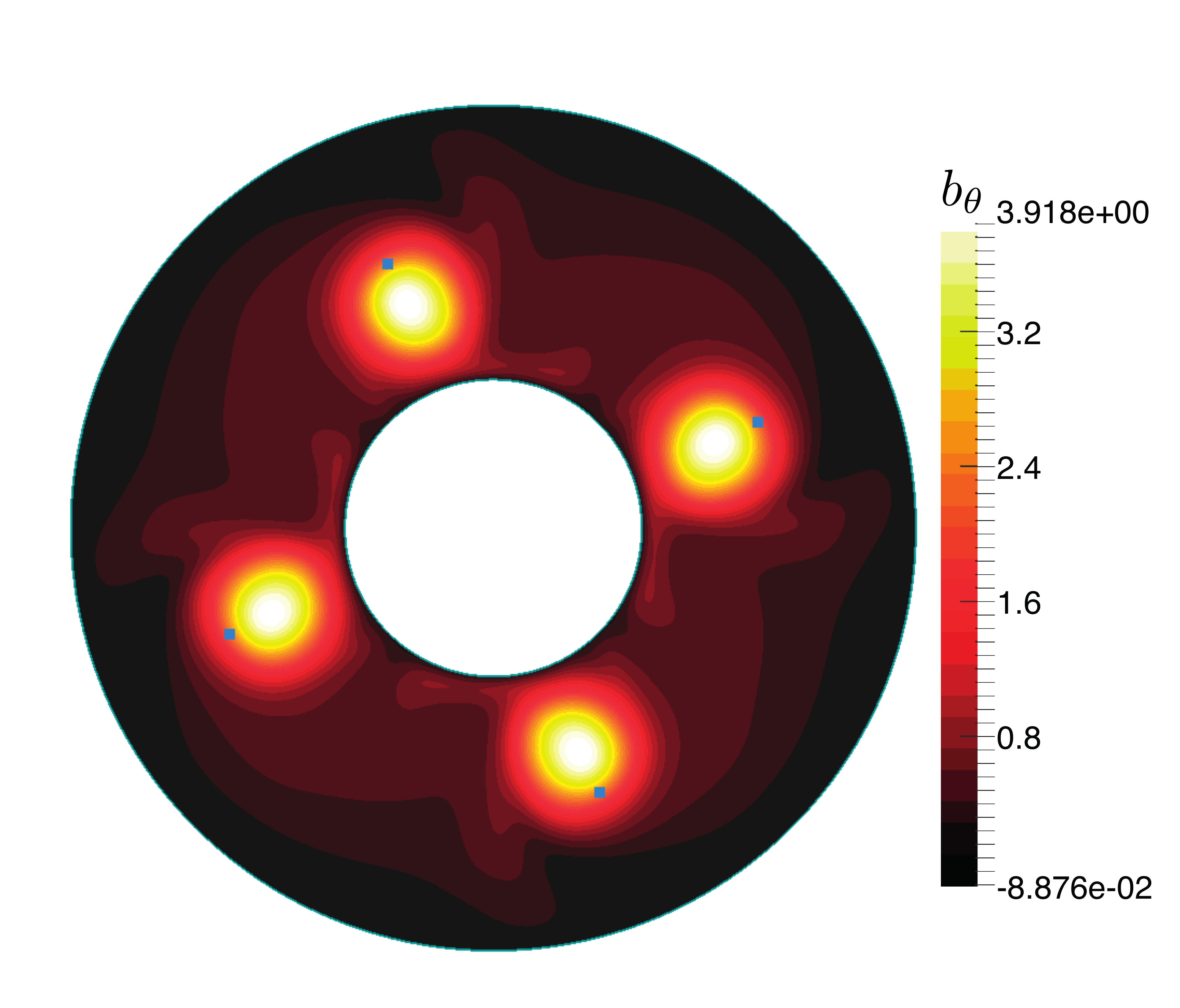}} &    \\  
      \setlength{\epsfysize}{7.0cm}
      \subfigure[]{\epsfbox{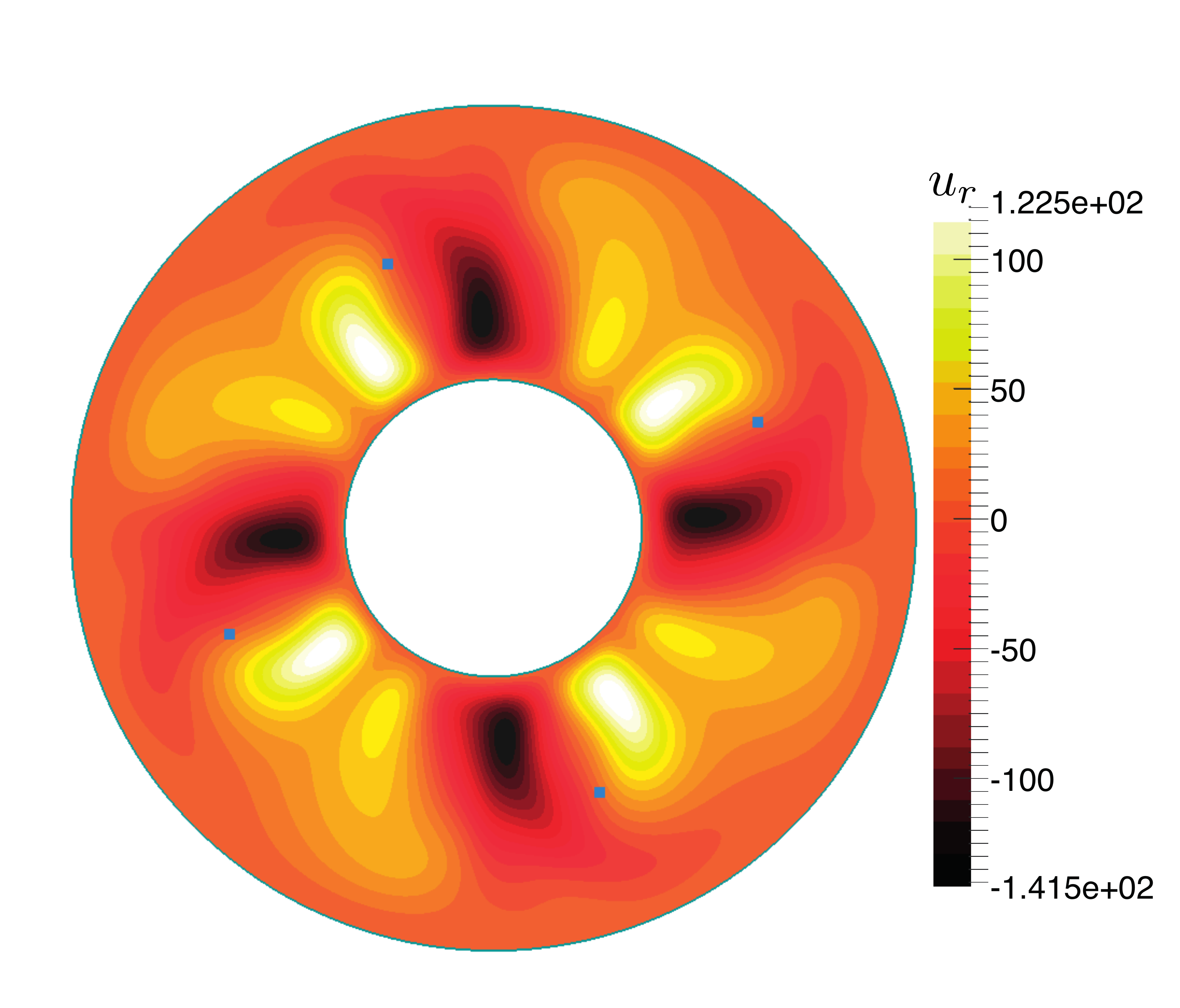}} &
      \setlength{\epsfysize}{7.0cm}
      \subfigure[]{\epsfbox{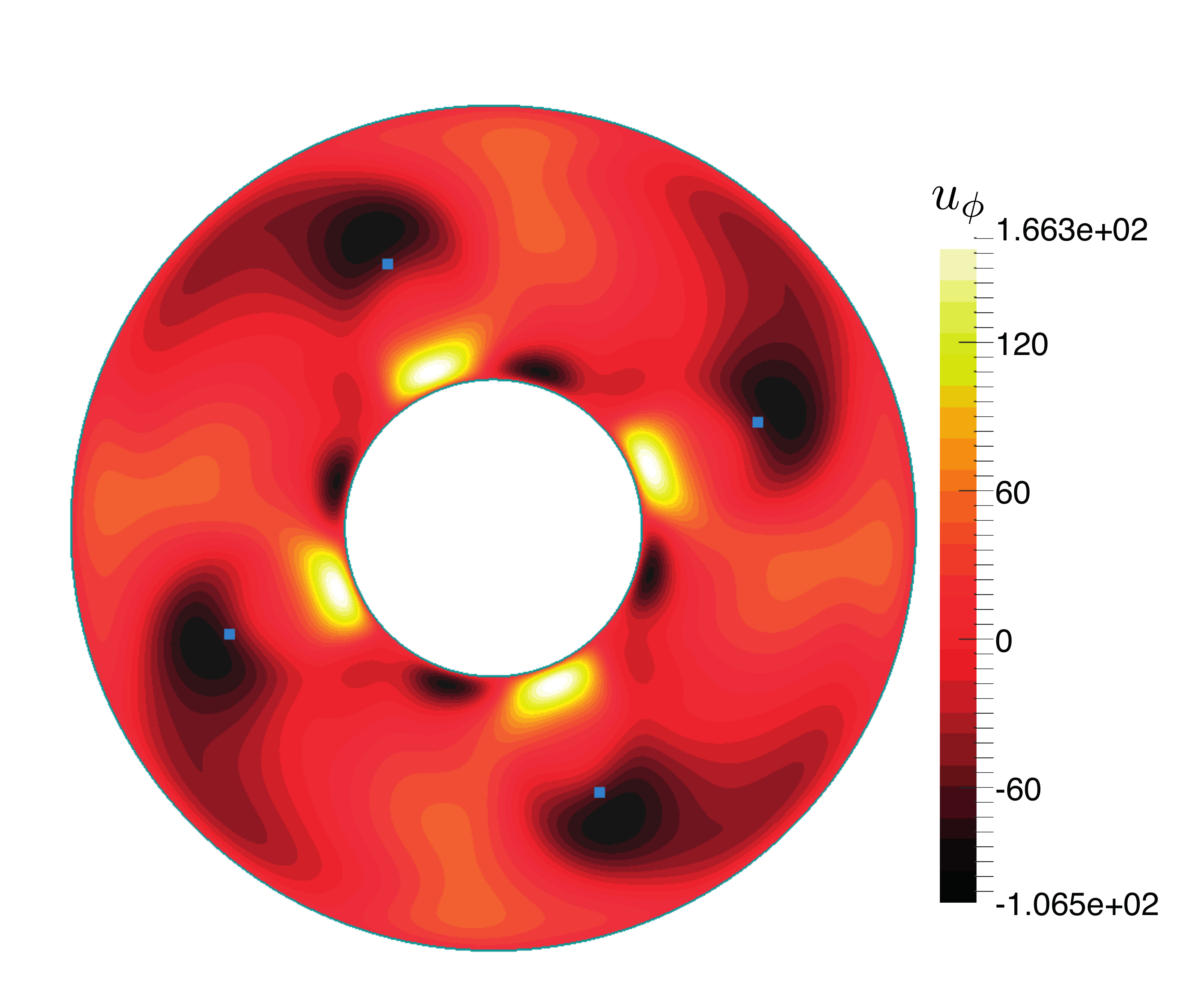}} 
    \end{tabular}
    \caption{Equatorial slices of the (steady) benchmark solution. The blue squares correspond to the reference points. (a) $\Theta$, (b) $b_\theta$, (c) $u_r$, (d) $u_{\phi}$.}
         \label{fig:eq_fv}                
  \end{center}
\end{figure}

\begin{figure}                 
  \begin{center}
  \setlength{\epsfysize}{5.50cm}
    \begin{tabular}{ccc}
      \setlength{\epsfysize}{5.50cm}
      \subfigure[]{\epsfbox{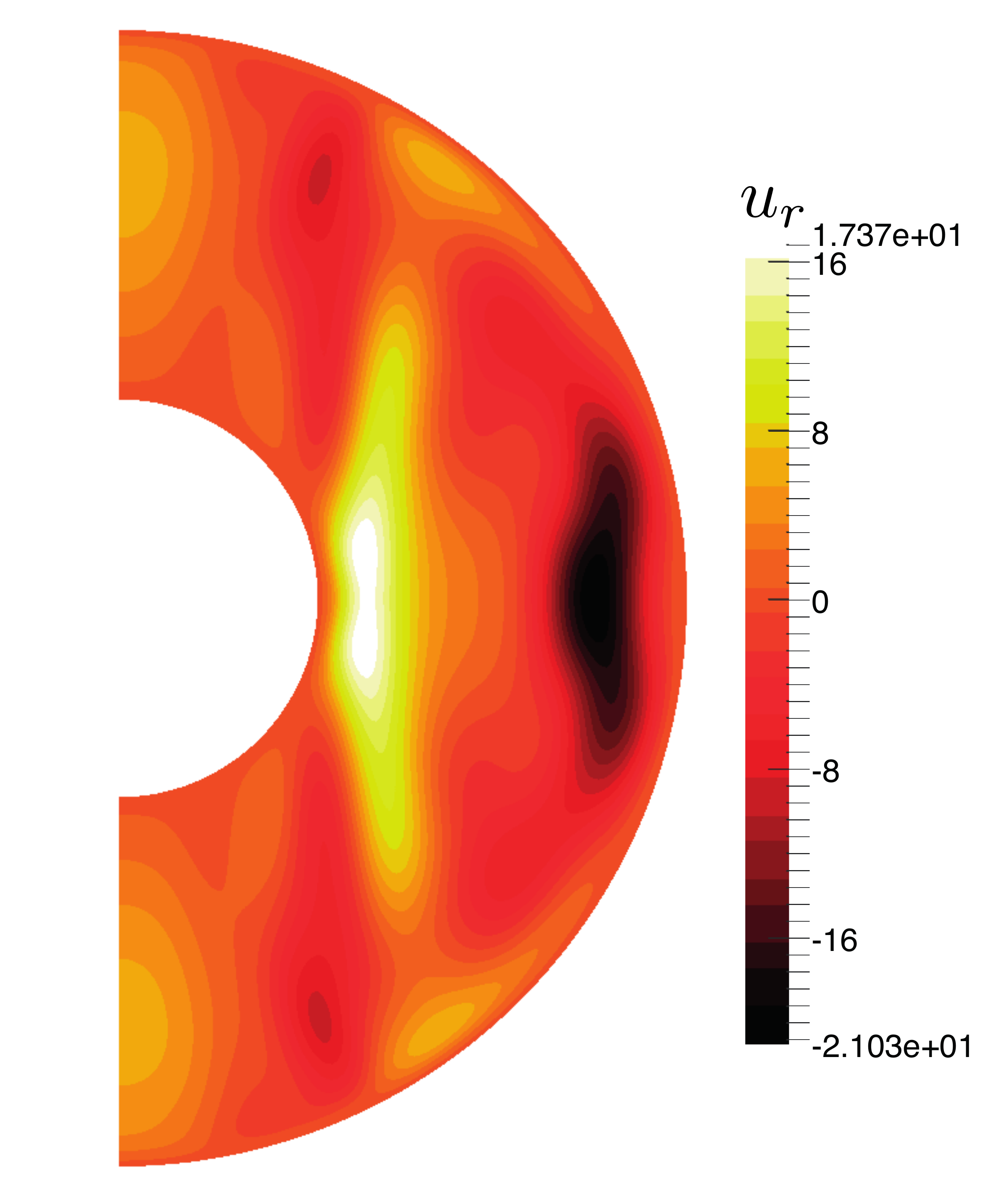}}  &
      \setlength{\epsfysize}{5.50cm}
      \subfigure[]{\epsfbox{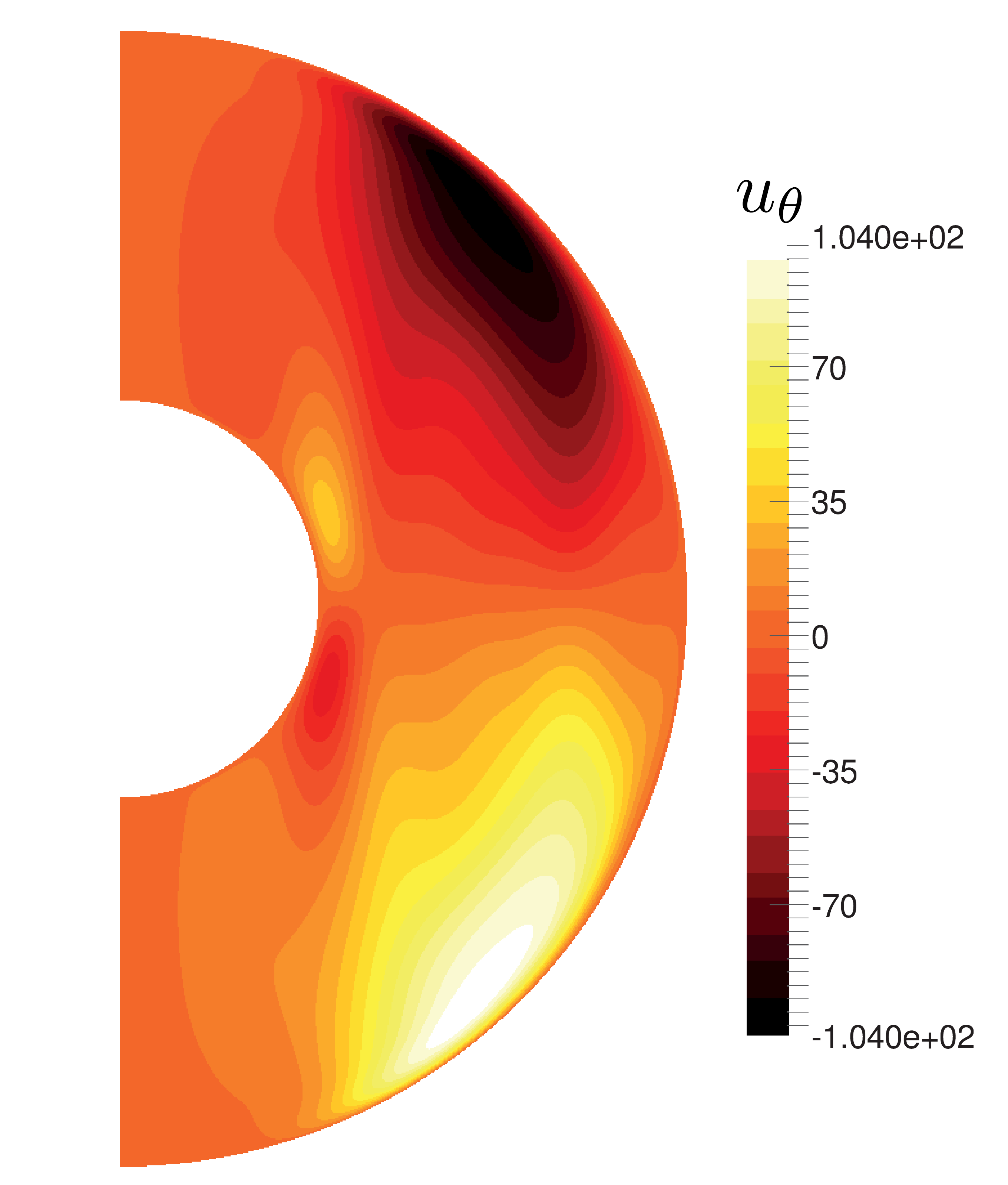}} &
      \setlength{\epsfysize}{5.50cm}
      \subfigure[]{\epsfbox{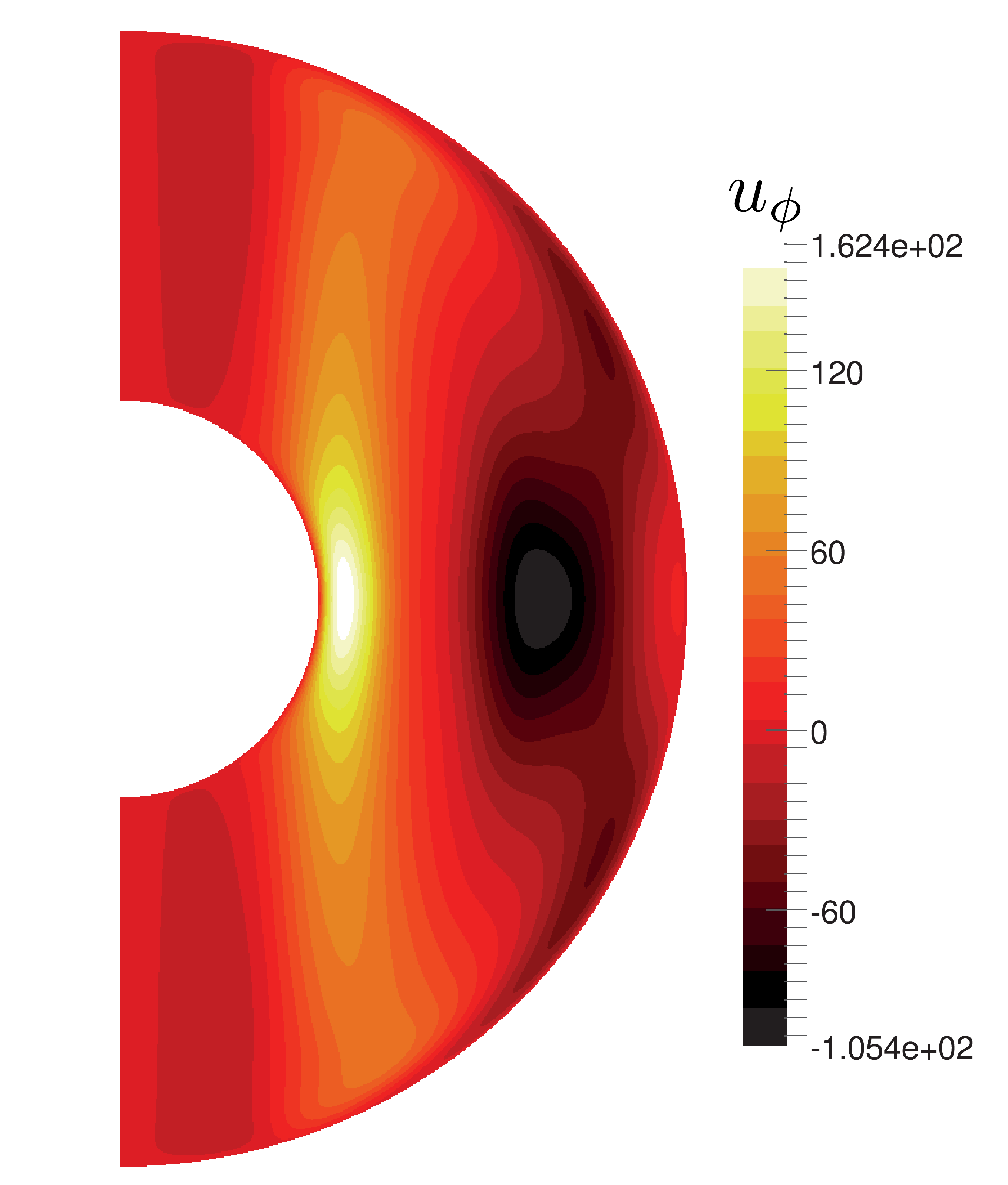}}  \\
            \setlength{\epsfysize}{5.50cm}
      \subfigure[]{\epsfbox{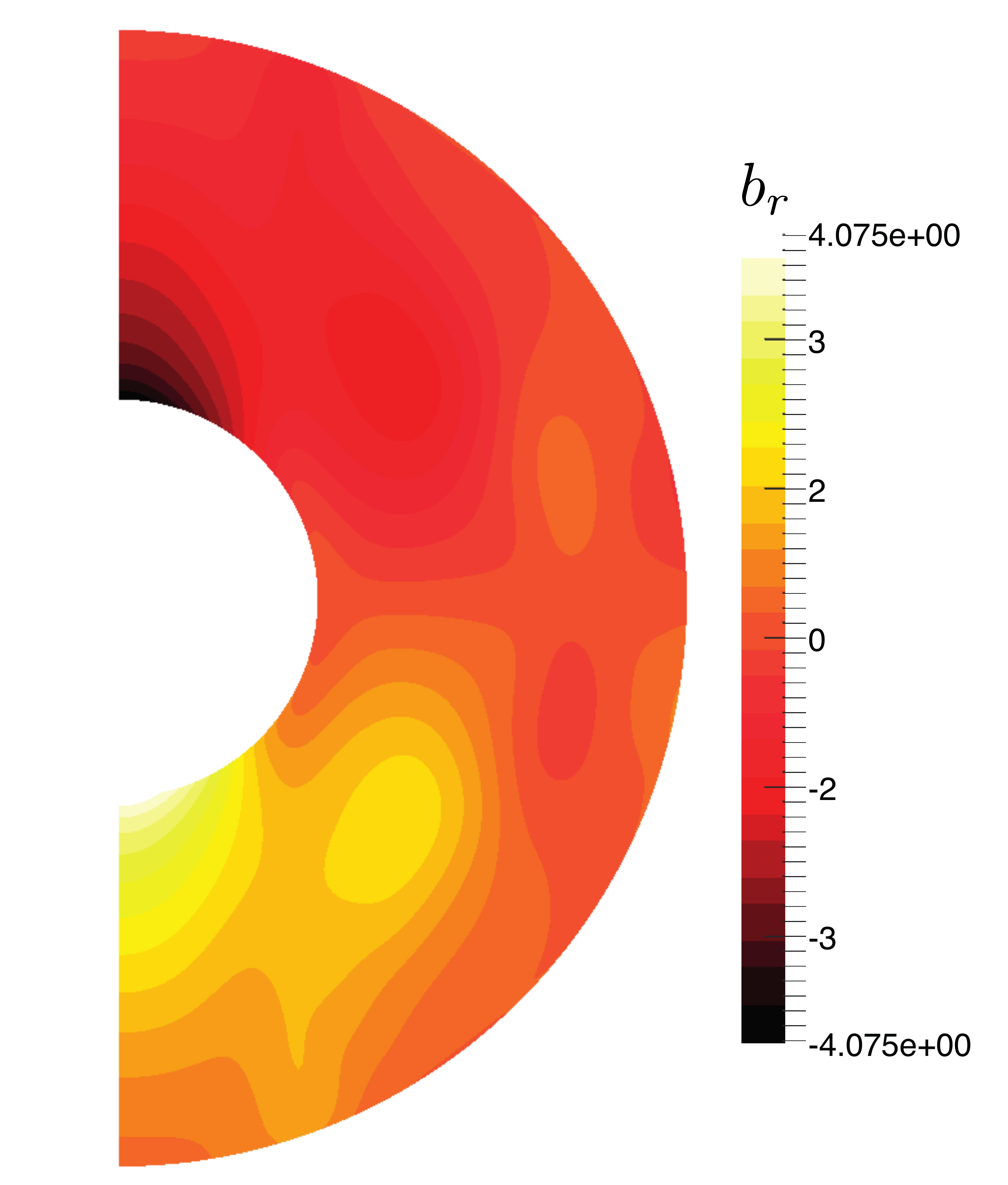}}  &
      \setlength{\epsfysize}{5.50cm}
      \subfigure[]{\epsfbox{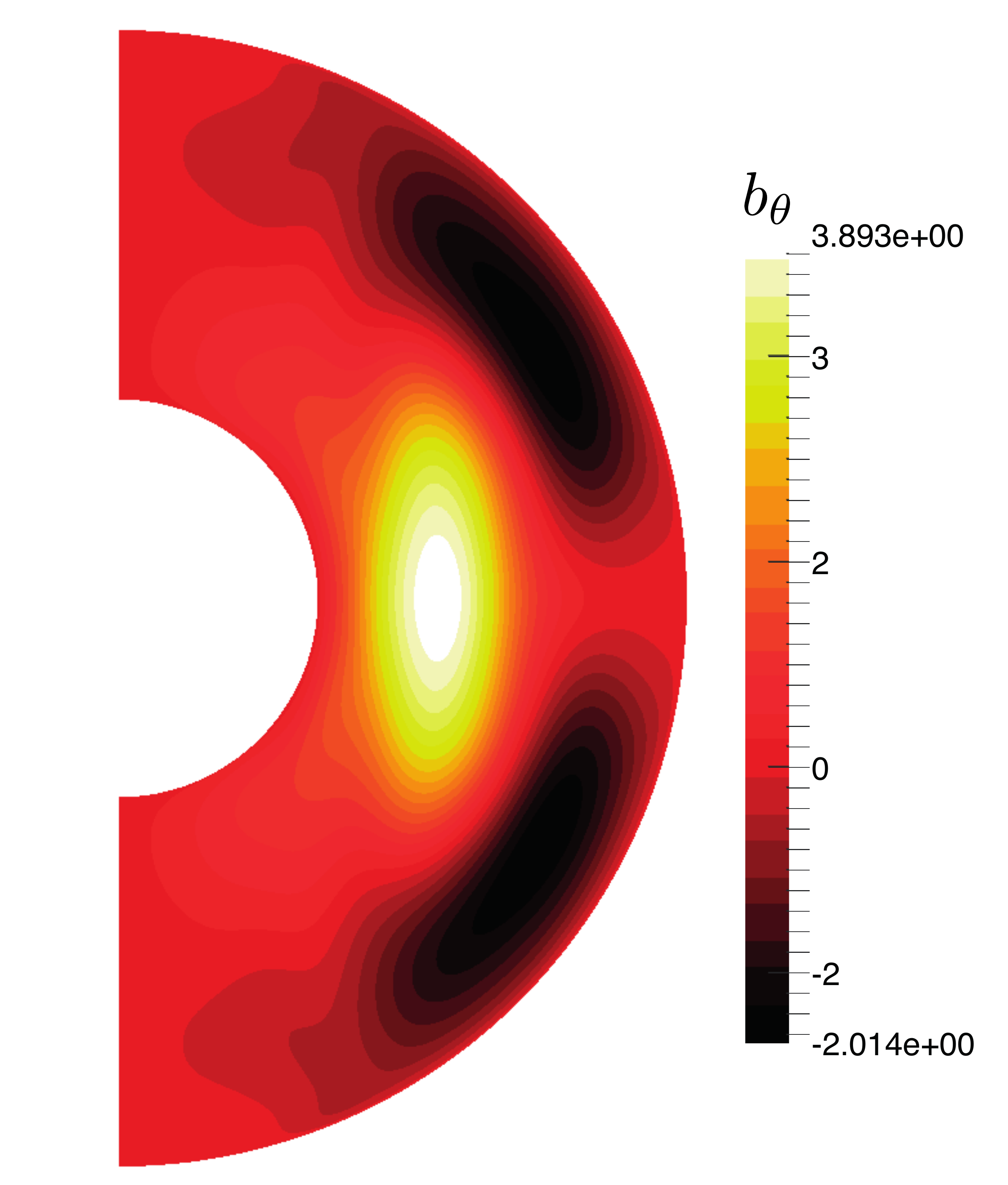}} &
      \setlength{\epsfysize}{5.50cm}
      \subfigure[]{\epsfbox{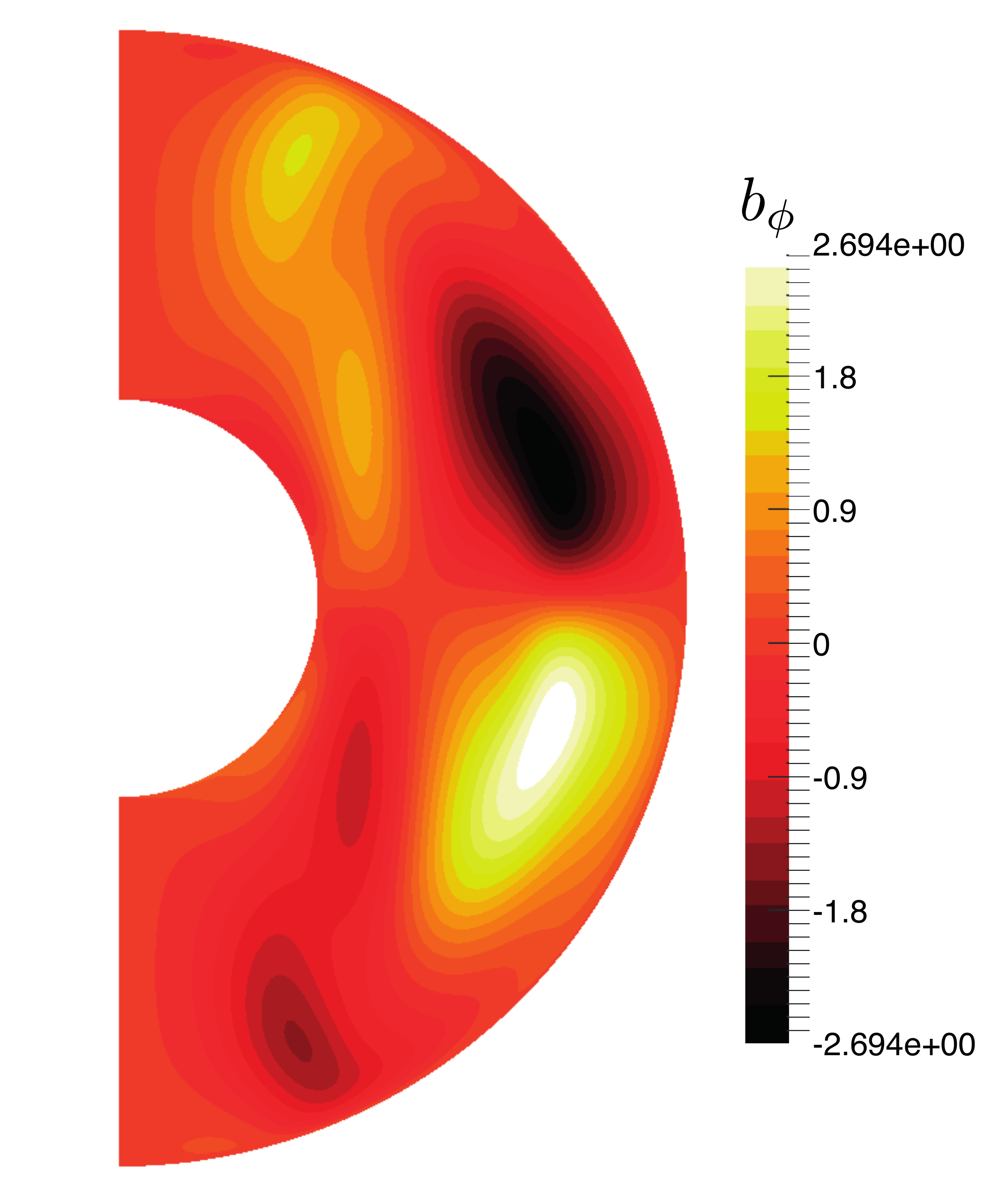}}  
    \end{tabular}
    \caption{Meridional slices of the steady benchmark solution in a plane that contains one of the reference points. (a) $u_r$, (b) $u_\theta$, (c) $u_\phi$, (d) $b_{r}$, (e) $b_\theta$, (f) $b_\phi$.}
         \label{fig:med_fv}                
  \end{center}
\end{figure}

We see that the FV simulations recover well the two branches of the dynamo solution. The maximum discrepancy between the solutions is about 3.5\% and 1\% for the oscillating and steadily solutions, respectively. The structure of the non-oscillatory solution is depicted in Figures \ref{fig:eq_fv} and \ref{fig:med_fv}; the meridional planes are chosen such that they contain one of the reference points, that is, one of the blue dots depicted in Figure \ref{fig:eq_fv}.

In the same spirit of \cite{christensen2001numerical} and \cite{jackson2014}, we suggest that  the quasi-steady solution can act as benchmark solution, as it allows the comparison of well-defined numbers. The final solution is reached within less than one magnetic diffusion time, whereas at least five magnetic diffusion times are required to reach the steady state in the benchmark by \citet{jackson2014}. Following the same predecessor studies, we report in Table \ref{tab:benchmark} the total kinetic and magnetic energy as global data and we also provide local data of $u_\phi,b_\theta$ and $\Theta$ for a point in the equatorial plane at mid-depth where $u_r=0$ and $\partial_\phi u_r > 0$.   
\begin{table}
\begin{center}
\begin{tabular}{c|ccc}
\hline \hline 
& FV64 & FV128 &PS\\ \hline
$E_{mag}$ & 309086.0(1.17\%) & 311950.2 (0.26\%) & 312754.7 \\
$E_{kin}$  & 21502.1 (0.61\%) & 21576.2 (0.27\%) & 21634.9 \\
$\Theta$ & 0.3920 (0.31\%) & 0.3925 (0.18\%) & 0.3932 \\
$B_{\theta}$ & 2.1510 (1.43\%)& 2.1839 (0.07\%) & 2.1823 \\
$V_{\phi}$ & -81.23 (0.36\%)& -80.74 (0.24\%) & -80.9318 \\
$\omega$ & 5.6453 (1.55\%) & 5.4959 (1.13\%) & 5.5588 \\
\hline
$\Delta t$& $4.0 \cdot 10^{-6}$ &$4.0 \cdot 10^{-6}$ & $9.2\cdot 10^{-6}$ \\
\hline
$\#$ cores & 64 & 25 &100  \\
CPU time/time step (sec)& 381 &3071 & 196\\
\hline
\hline
\end{tabular}
\caption{Benchmark values for the steadily drifting dynamo at $q=8$. FV64 and FV128 refer to results obtained with the FV code with six blocks of $64^3$ and $128^3$ grid points respectively. PS denotes pseudospectral solution. The pseudospectral results were obtained using 200 grid points in radius and a spherical harmonic truncation of $L=100,M=100$. The numbers between brackets give the difference between the respective finite-volume and pseudospectral solution. Also shown are the time integration step $\Delta t$ and a typical CPU time required to perform one time integration step.}
\label{tab:benchmark}
\end{center}
\end{table}

We find that the results of the FV code are in good agreement with the spectral ones; the discrepancies for the magnetic `quantities' are considerably smaller than the ones reported in \cite{jackson2014}. 

\section{Conclusions}
In this article we have described the implementation and validation of an unstructured finite-volume code for the solution of the incompressible MHD equations. The main novelty of this numerical tool, is the capability of working with unstructured meshes, which allows for a much broader range of geometries to be considered. The examples discussed in this work were mainly motivated by the study of natural dynamos. Our method, however, is also capable of modeling the complex geometries that are encountered in many dynamo experiments, such as geometries that incorporate the impellers in the VKS and Madison experiment \citep{monchaux2007generation,nornberg2006} or the blades in the Riga experiment \citep{gailitis2000detection}. This versatility is the main strength of the FV method. For spherical geometries, however, pseudo-spectral methods remain the \textit{nec plus ultra} in terms of convergence and computational efficiency.

Furthermore, we have presented a new benchmark case of a thermally driven self-consistent dynamo simulation. This complements earlier works of \cite{christensen2001numerical,jackson2014} and \cite{marti2014}. Given that it uses ferromagnetic boundary conditions, it is particularly attractive for testing local codes; a further advantage is that a steadily drifting solution is obtained after less than two decay times, in contrast to the benchmark of \cite{jackson2014}.

The main challenge remaining is the implementation of insulating boundary conditions, which are more relevant in the context of planetary physics. 
   
\section*{Acknowledgements}
Funding for this work from the ERC grant 247303 `MFECE' and the SNF grant 200020 143596 is gratefully acknowledged. This work was also supported by a grant from the Swiss National Supercomputing Centre (CSCS) under project IDs s225 and s369 for which we are grateful. We thank D. C\'ebron for providing some of the results displayed in Table \ref{tab:decay_geometries}. Finally, we thank two anonymous reviewers for their constructive comments that helped to improve this article. 

\appendix
\section{Stability properties of the mixed Adams-Bashforth/Crank-Nicolson scheme} \label{app:stab}
In this appendix, we will demonstrate that the mixed Adams-Bashforth/Crank-Nicolson scheme for the nonlinear terms in the MHD equations is stable, independent of the time step $\Delta t$. This is in fact an extension of the work of \cite{ham2007} who showed that this scheme was stable for the incompressible Navier-Stokes equations. As a starting point, we recapitulate their line of thought. It will be useful to first introduce the kinetic energy norm $\sum_{i} V_i {\boldsymbol U}_i^2$. 
The advective term in the Navier-Stokes equations under the mixed Adams-Bashforth/Crank-Nicolson scheme reads: 
\begin{equation}
V_i\left(\mathbb{C}_U {\boldsymbol U} \right)_i = \sum_{j \in \pi_i} \frac{{\boldsymbol U}_i^{CN} + {\boldsymbol U}_j^{CN}}{2}U_{ij}^{AB}, 
\end{equation}
where the superscripts $AB$ and $CN$ denote a second-order Adams-Bashforth and Crank-Nicolson time discretisation, respectively. We can cast this in the following matrix representation:
\begin{equation}\frac{1}{2}
\left(
\begin{array}{ccccc}
 0 & ... & ... & ... & ... \\
 ... & 0  & ... & U_{ij}^{AB} & ... \\
  ... & ...  & 0 & ... & ... \\
   ... & U_{ji}^{AB}  & ... & 0 & ... \\ 
   .. & ...  & ... & ... & 0 
\end{array}
\right)
\left(
\begin{array}{c}
\vdots \\ {\boldsymbol U}_i^{CN} \\ \vdots \\ {\boldsymbol U}_j^{CN} \\ \vdots
\end{array}
\right)
\end{equation}
The essential point now is that  the coefficient matrix is skew-symmetric, This can be seen as follows. According to eq. (\ref{incompress-np1}), the diagonal entries are $\frac{1}{2}\sum_{j \in \pi_{i}} U_{ij}^{AB}=0$ given that $\sum_{j \in \pi_{i}} U_{ij}^{n}=0$ and $\sum_{j \in \pi_{i}} U_{ij}^{n-1}=0$ by construction. Furthermore, the fluxes satisfy $U_{ij}=-U_{ji}$ by definition, that is, $\mathbb{C_U} = -\mathbb{C}_U$. 

We can now write the time-advanced scheme \textcolor{red}{in} in the following form:
\begin{equation}
V_i\frac{{\boldsymbol U}^{\star}_i-{\boldsymbol U}^n_i}{\Delta t } = -V_i\left(\mathbb{C}_U {\boldsymbol U}^{CN} \right)_{i} = -\left(\mathbb{\tilde C}_U {\boldsymbol U}^{CN} \right)_{i}  .
\end{equation}
We now take the dot-product of the above expression with $({\boldsymbol U}^{\star}_i + {\boldsymbol U}^{n}_i)$ and take the sum over all CVs. This yields:
\begin{equation}
 \sum_{i} V_i ({\boldsymbol U}^{\star}_i+{\boldsymbol U}^n_i) \cdot \frac{{\boldsymbol U}^{\star}-{\boldsymbol U}^n}{\Delta t } = -\frac{1}{2}\sum_i ({\boldsymbol U}^{\star}+{\boldsymbol U}^n)_i  \cdot \left[ \mathbb{\tilde C}_U ({\boldsymbol U}^{\star}+{\boldsymbol U}^n) \right]_i. 
 \end{equation} 
 Due to the skew-symmetric character of the operator (or matrix) $\mathbb{\tilde C}_{U}$ the right-hand side of this expression vanishes, and this implies that the kinetic energy norm is conserved. It follows that the mixed Adams-Bashforth/Crank-Nicolson scheme is stable for the Navier-Stokes equation.
 
This  can now be easily extended to the MHD case. To this end, we consider the `MHD' energy norm $\sum_{i} V_i \left({\boldsymbol U}^2_i + {\boldsymbol B}^2_{i} \right)$. The time integration schemes for the nonlinear terms in the MHD equations can be written as follows:
\begin{equation}
V_i\frac{{\boldsymbol U}^{\star}_i-{\boldsymbol U}^n_i}{\Delta t } = -V_i \left(\mathbb{C}_U {\boldsymbol U}^{CN} \right)_{i} + V_i \left(\mathbb{C}_B {\boldsymbol B}^{CN} \right)_{i} = -\left(\mathbb{\tilde C}_U {\boldsymbol U}^{CN} \right)_{i} +  \left(\mathbb{\tilde C}_B {\boldsymbol B}^{CN} \right)_{i} ,
\end{equation}
\begin{equation}
V_i\frac{{\boldsymbol B}^{\star}_i-{\boldsymbol B}^n_i}{\Delta t } = -V_i \left(\mathbb{C}_U {\boldsymbol B}^{CN} \right)_{i} + V_i \left(\mathbb{C}_B {\boldsymbol U}^{CN} \right)_{i} = - \left(\mathbb{\tilde C}_U {\boldsymbol B}^{CN} \right)_{i} + \left(\mathbb{\tilde C}_B {\boldsymbol U}^{CN} \right)_{i} .
\end{equation}
After some algebra, we find that
\begin{multline}
\frac{1}{\Delta t}\sum_{i} V_i \left\{ \left({\boldsymbol U}^{\star}_i\right)^2+\left({\boldsymbol B}^{\star}_i\right)^2- \left({\boldsymbol U}^{n}_i\right)^2-\left({\boldsymbol B}^{n}_i\right)^2 \right\}\\  =  \frac{1}{2}\sum_{i }\left\{ -  ( {\boldsymbol U}^{\star}+{\boldsymbol U}^n)_i  \cdot \left[ \mathbb{ \tilde C}_U ({\boldsymbol U}^{\star}+{\boldsymbol U}^n) \right]_i  -  ( {\boldsymbol B}^{\star}+{\boldsymbol B}^n)_i  \cdot \left[ \mathbb{ \tilde C}_U ({\boldsymbol B}^{\star}+{\boldsymbol B}^n) \right]_i  \right. \\ \left. + ( {\boldsymbol U}^{\star}+{\boldsymbol U}^n)_i  \cdot \left[ \mathbb{ \tilde C}_B ({\boldsymbol B}^{\star}+{\boldsymbol B}^n) \right]_i  +  ( {\boldsymbol B}^{\star}+{\boldsymbol B}^n)_i  \cdot \left[ \mathbb{ \tilde C}_B ({\boldsymbol U}^{\star}+{\boldsymbol U}^n) \right]_i  \right\}.
\end{multline}
Following similar arguments as those outlined above, the first two terms on the right-hand side of this expression are both strictly zero because of the skew-symmetric nature of $\mathbb{\tilde C}_{U}$. Likewise, $\mathbb{\tilde C}_B$ is also skew-symmetric and therefore, the last two terms on the right-hand side cancel each other. Thus, we find that the `MHD energy norm' $\sum_{i} V_i({\boldsymbol U}_i^2 + {\boldsymbol B}_i^2)$ is conserved. This implies that no unbounded growth of the total energy can occur, and we can conclude that the mixed Adans-Bashforth/Crank-Nicolson scheme is stable.

\section{Analytical solution of the decay modes in a full sphere and a spherical shell with pseudo-vacuum boundary conditions} \label{app:decay}
We consider a toroidal-poloidal decomposition of the magnetic field:
\begin{equation}
{\boldsymbol b} = \nabla \times T{\boldsymbol r} + \nabla \times \nabla \times S{\boldsymbol r}.  
\end{equation}
It is customary to expand $T$ and $S$ in a series of spherical harmonics:
\begin{equation}
T = \sum_{l=0}^{\infty}\sum_{m=-l}^{l} T_{lm}(r) Y_{lm}(\theta, \phi), S = \sum_{l=0}^{\infty}\sum_{m=-l}^{l} S_{lm}(r) Y_{lm}(\theta, \phi),
\end{equation}
where the functions  $Y_{lm}$ denote the spherical harmonics of degree $l$ and order $m$. 

Using this decomposition, we can recast the magnetic diffusion equation 
\begin{equation}
\frac{\partial {\boldsymbol b}}{\partial t} = \nabla^2 {\boldsymbol b},
\end{equation}
as a set of differential equations for $T_{lm}$ and $S_{lm}$:
\begin{equation}
\frac{\partial T_{lm}}{\partial t} = \frac{1}{r^2}\frac{\partial}{\partial r} \left(r^2 \frac{\partial T_{lm}}{\partial r} \right) - \frac{l(l+1)}{r^2}T_{lm}, \label{eq:tor_eq}
\end{equation}
\begin{equation}
 \frac{\partial S_{lm}}{\partial t} = \frac{1}{r^2}\frac{\partial}{\partial r} \left(r^2 \frac{\partial S_{lm}}{\partial r} \right) - \frac{l(l+1)}{r^2}S_{lm}.  \label{eq:pol_eq}
\end{equation}
The pseudo-vacuum boundary condition (\ref{eq:bcb}) in terms of the functions $T_{lm}$ and $S_{lm}$ is the following:
\begin{equation}
T_{lm} = 0,  \label{eq:tor_bc}
\end{equation}
\begin{equation}
\frac{\partial(r S_{lm})}{\partial r} = 0.   \label{eq:pol_bc}
\end{equation}
We now look for solutions of (\ref{eq:tor_eq})-(\ref{eq:tor_bc}) and (\ref{eq:pol_eq})-(\ref{eq:pol_bc}) of the form $T_{lm},S_{lm} = f(r)\exp(-\sigma t)$. It follows that $T_{lm},S_{lm}$ are of the form:
\begin{equation}
T_{lm},S_{lm} = \alpha j_{l}(kr) + \beta y_l(kr),
\end{equation}
where $\sigma=k^2$ and $j_{l}$ and $y_{l}$ refer to the spherical Bessel and Neumann functions, respectively. In the case of a full sphere geometry, the coefficient $\beta$ should be zero so that the solution remains regular at the origin $r=0$. The boundary condition at $r=1$ then quantizes the possible values for $k$, and thus for the decay rates $\sigma$. The slowest toroidal decay rate corresponds to the lowest zero of $j_l$; this gives $k=4.493$ and $\sigma=20.19$. Similary, we find $\sigma=7.5279$ for the slowest decaying poloidal eigenmode. 

The eigenvalues and eigenmodes for a spherical shell geometry ($r_i=7/13, r_o=20/13$) are found by imposing the boundary conditions (\ref{eq:tor_bc})-(\ref{eq:pol_bc}) at $r=r_i$ and $r=r_o$. For the toroidal mode this gives the conditions:
\begin{eqnarray}
\alpha j_l(kr_i) + \beta y_l(kr_i) & = & 0, \\
\alpha j_l(kr_o) + \beta y_l(kr_o) & = & 0.
\end{eqnarray}
This homogeneous system of two equations for $\alpha,\beta$ has only non-trivial solutions if the determinant of the coefficient matrix is zero, that is, if $j_l(kr_i) y_l(kr_o)- y_l(kr_i) j_l(kr_o)=0$. The roots of this transcendent equation yield admissible values for $k$ (and thus $\sigma$). The lowest zero is $k=3.261$, and the toroidal eigenmode with the slowest decay has a decay rate $\sigma=k^2=10.634$. Along similar lines, we find $\sigma=2.227904$ for the poloidal mode.
\bibliographystyle{gji}
\nocite{*}
\bibliography{GJI}
\end{document}